\documentclass[12pt,letterpaper]{article}
\usepackage[utf8]{inputenc}
\usepackage[T1]{fontenc}
\usepackage{amsmath,amsfonts,amssymb,amsthm}
\usepackage{graphicx,geometry,parskip,placeins,enumitem}
\usepackage{xcolor}
\usepackage[authoryear]{natbib}
\definecolor{linkcolor}{RGB}{0,90,160}
\usepackage[colorlinks=true,linkcolor=linkcolor,urlcolor=linkcolor,citecolor=linkcolor]{hyperref}
\newcommand{\E}{\mathbb{E}}
\renewcommand{\P}{\mathbb{P}}

\providecommand{\ChicagoClock}{4:21}
\providecommand{\ChicagoDate}{November 5, 2019}
\providecommand{\ChicagoFinalAwayScore}{118}
\providecommand{\ChicagoFinalHomeScore}{112}

\providecommand{\ChicagoMaxWP}{96.9\%}
\providecommand{\ChicagoMinutesRemaining}{16.4}
\providecommand{\ChicagoPIT}{0.968}
\providecommand{\ChicagoPeakAwayScore}{67}
\providecommand{\ChicagoPeakHomeScore}{85}
\providecommand{\ChicagoPeakLead}{18}

\providecommand{\ChicagoStartingWP}{71.2\%}
\providecommand{\ChicagoTailProbability}{3.2\%}

\providecommand{\DyadicBootstrapReplicates}{9,999}
\providecommand{\DyadicBootstrapSeed}{20260815}
\providecommand{\NBACompletedGames}{8,290}
\providecommand{\NBACoverageRows}{
2018 & 1,230 & 1,230 & 1,222 & 8 \\
2019 & 1,230 & 1,230 & 1,230 & 0 \\
2020 & 1,062 & 1,059 & 1,053 & 6 \\
2021 & 1,112 & 1,080 & 1,080 & 0 \\
2022 & 1,241 & 1,230 & 1,230 & 0 \\
2023 & 1,231 & 1,230 & 1,230 & 0 \\
2024 & 1,233 & 1,231 & 1,231 & 0 \\
}
\providecommand{\NBACrossingNinetyCILower}{2.7\%}
\providecommand{\NBACrossingNinetyCIUpper}{4.7\%}
\providecommand{\NBACrossingNinetyEpisodes}{9,371}
\providecommand{\NBACrossingNinetyFiveAllCILower}{1.4\%}
\providecommand{\NBACrossingNinetyFiveAllCIUpper}{3.1\%}

\providecommand{\NBACrossingNinetyFiveAllResidual}{2.3\%}
\providecommand{\NBACrossingNinetyFiveCILower}{1.4\%}
\providecommand{\NBACrossingNinetyFiveCIUpper}{3.2\%}
\providecommand{\NBACrossingNinetyFiveEpisodes}{8,663}
\providecommand{\NBACrossingNinetyFiveGames}{8,156}
\providecommand{\NBACrossingNinetyFiveImplied}{4.2\%}
\providecommand{\NBACrossingNinetyFiveLoss}{6.5\%}
\providecommand{\NBACrossingNinetyFiveResidual}{2.3\%}
\providecommand{\NBACrossingNinetyFiveStrictCILower}{1.5\%}
\providecommand{\NBACrossingNinetyFiveStrictCIUpper}{3.2\%}

\providecommand{\NBACrossingNinetyFiveStrictResidual}{2.3\%}
\providecommand{\NBACrossingNinetyGames}{8,250}
\providecommand{\NBACrossingNinetyImplied}{8.8\%}
\providecommand{\NBACrossingNinetyLoss}{12.5\%}
\providecommand{\NBACrossingNinetyNineCILower}{0.1\%}
\providecommand{\NBACrossingNinetyNineCIUpper}{0.9\%}
\providecommand{\NBACrossingNinetyNineEpisodes}{7,660}
\providecommand{\NBACrossingNinetyNineGames}{7,598}
\providecommand{\NBACrossingNinetyNineImplied}{0.8\%}
\providecommand{\NBACrossingNinetyNineLoss}{1.3\%}
\providecommand{\NBACrossingNinetyNineResidual}{0.5\%}
\providecommand{\NBACrossingNinetyResidual}{3.7\%}
\providecommand{\NBADyadicCalendarGlobalD}{0.097}
\providecommand{\NBADyadicCalendarGlobalP}{\ensuremath{<0.001}}
\providecommand{\NBAEligibleGames}{8,290}
\providecommand{\NBAExcludePatchedGlobalD}{0.096}
\providecommand{\NBAExcludePatchedGlobalP}{\ensuremath{<0.001}}
\providecommand{\NBAExcludedExhibitionEvents}{10}
\providecommand{\NBAExcludedExhibitionFeeds}{5}
\providecommand{\NBAFixedClockExcluded}{11}
\providecommand{\NBAFixedClockGames}{8,265}
\providecommand{\NBAFixedClockLOCFCells}{195}
\providecommand{\NBAFixedClockLOCFRMS}{2.8\%}
\providecommand{\NBAFixedClockLOCFSignificantCells}{3}
\providecommand{\NBAFixedClockLinearCells}{196}
\providecommand{\NBAFixedClockLinearRMS}{2.7\%}
\providecommand{\NBAFixedClockLinearSignificantCells}{2}
\providecommand{\NBAFranchiseLinkedGlobalD}{0.097}
\providecommand{\NBAFranchiseLinkedGlobalP}{\ensuremath{<0.001}}
\providecommand{\NBAGames}{8,276}

\providecommand{\NBAGlobalD}{0.097}
\providecommand{\NBAGlobalDyadicP}{\ensuremath{<0.001}}
\providecommand{\NBALOSODMax}{0.102}
\providecommand{\NBALOSODMin}{0.093}
\providecommand{\NBALOSOTailMax}{1.5\%}
\providecommand{\NBALOSOTailMin}{1.2\%}
\providecommand{\NBALeaveSeasonRows}{
2018 & 7,054 & 0.093 & 1.5\% \\
2019 & 7,046 & 0.097 & 1.2\% \\
2020 & 7,223 & 0.094 & 1.3\% \\
2021 & 7,196 & 0.100 & 1.3\% \\
2022 & 7,046 & 0.102 & 1.3\% \\
2023 & 7,046 & 0.093 & 1.2\% \\
2024 & 7,045 & 0.101 & 1.5\% \\
}
\providecommand{\NBAMedianUpdatesPerGame}{466}
\providecommand{\NBAMinimumFeedCoverage}{99.3\%}

\providecommand{\NBAMissingFeedGames}{14}

\providecommand{\NBAPatchedGames}{16}
\providecommand{\NBAPublishedUpdates}{3,877,820}
\providecommand{\NBARawGlobalD}{0.097}
\providecommand{\NBARawGlobalP}{\ensuremath{<0.001}}
\providecommand{\NBARoundingLowerGlobalD}{0.095}

\providecommand{\NBARoundingLowerTailNinetyFiveRate}{6.3\%}
\providecommand{\NBARoundingUpperGlobalD}{0.099}

\providecommand{\NBARoundingUpperTailNinetyFiveRate}{6.6\%}
\providecommand{\NBAScheduledGames}{8,339}
\providecommand{\NBATailNinetyCILower}{1.4\%}
\providecommand{\NBATailNinetyCIUpper}{4.0\%}
\providecommand{\NBATailNinetyExcess}{2.7\%}
\providecommand{\NBATailNinetyFiveAtomRate}{0.00\%}
\providecommand{\NBATailNinetyFiveCILower}{0.4\%}
\providecommand{\NBATailNinetyFiveCIUpper}{2.3\%}
\providecommand{\NBATailNinetyFiveExcess}{1.3\%}
\providecommand{\NBATailNinetyFiveRate}{6.3\%}
\providecommand{\NBATailNinetyFiveStrictRate}{6.3\%}
\providecommand{\NBATailNinetyNineCILower}{-0.3\%}
\providecommand{\NBATailNinetyNineCIUpper}{0.5\%}
\providecommand{\NBATailNinetyNineExcess}{0.1\%}
\providecommand{\NBATailNinetyNineRate}{1.1\%}
\providecommand{\NBATailNinetyRate}{12.7\%}

\providecommand{\NBAUpdateQOne}{447}
\providecommand{\NBAUpdateQThree}{488}
\providecommand{\NFLCompletedGames}{1,855}
\providecommand{\NFLCoverageRows}{
2018 & 256 & 254 & 254 & 0 \\
2019 & 256 & 255 & 255 & 0 \\
2020 & 256 & 255 & 255 & 0 \\
2021 & 272 & 271 & 271 & 0 \\
2022 & 272 & 269 & 269 & 0 \\
2023 & 272 & 272 & 272 & 0 \\
2024 & 272 & 272 & 272 & 0 \\
}
\providecommand{\NFLCrossingNinetyCILower}{-0.9\%}
\providecommand{\NFLCrossingNinetyCIUpper}{2.8\%}
\providecommand{\NFLCrossingNinetyEpisodes}{1,974}
\providecommand{\NFLCrossingNinetyFiveAllCILower}{-0.2\%}
\providecommand{\NFLCrossingNinetyFiveAllCIUpper}{2.7\%}

\providecommand{\NFLCrossingNinetyFiveAllResidual}{1.2\%}
\providecommand{\NFLCrossingNinetyFiveCILower}{-0.2\%}
\providecommand{\NFLCrossingNinetyFiveCIUpper}{2.9\%}
\providecommand{\NFLCrossingNinetyFiveEpisodes}{1,842}
\providecommand{\NFLCrossingNinetyFiveGames}{1,772}
\providecommand{\NFLCrossingNinetyFiveImplied}{3.3\%}
\providecommand{\NFLCrossingNinetyFiveLoss}{4.6\%}
\providecommand{\NFLCrossingNinetyFiveResidual}{1.3\%}
\providecommand{\NFLCrossingNinetyFiveStrictCILower}{-0.2\%}
\providecommand{\NFLCrossingNinetyFiveStrictCIUpper}{2.9\%}

\providecommand{\NFLCrossingNinetyFiveStrictResidual}{1.3\%}
\providecommand{\NFLCrossingNinetyGames}{1,818}
\providecommand{\NFLCrossingNinetyImplied}{7.6\%}
\providecommand{\NFLCrossingNinetyLoss}{8.5\%}
\providecommand{\NFLCrossingNinetyNineCILower}{0.2\%}
\providecommand{\NFLCrossingNinetyNineCIUpper}{2.2\%}
\providecommand{\NFLCrossingNinetyNineEpisodes}{1,705}
\providecommand{\NFLCrossingNinetyNineGames}{1,686}
\providecommand{\NFLCrossingNinetyNineImplied}{0.5\%}
\providecommand{\NFLCrossingNinetyNineLoss}{1.6\%}
\providecommand{\NFLCrossingNinetyNineResidual}{1.1\%}
\providecommand{\NFLCrossingNinetyResidual}{0.9\%}
\providecommand{\NFLDyadicCalendarGlobalD}{0.015}
\providecommand{\NFLDyadicCalendarGlobalP}{\ensuremath{0.856}}
\providecommand{\NFLEligibleGames}{1,848}
\providecommand{\NFLExcludePatchedGlobalD}{0.014}
\providecommand{\NFLExcludePatchedGlobalP}{\ensuremath{0.769}}

\providecommand{\NFLFixedClockGames}{1,848}
\providecommand{\NFLFixedClockLOCFCells}{197}
\providecommand{\NFLFixedClockLOCFRMS}{2.6\%}
\providecommand{\NFLFixedClockLOCFSignificantCells}{8}
\providecommand{\NFLFixedClockLinearCells}{197}
\providecommand{\NFLFixedClockLinearRMS}{2.9\%}
\providecommand{\NFLFixedClockLinearSignificantCells}{8}
\providecommand{\NFLFranchiseLinkedGlobalD}{0.015}
\providecommand{\NFLFranchiseLinkedGlobalP}{\ensuremath{0.705}}
\providecommand{\NFLGames}{1,848}

\providecommand{\NFLGlobalD}{0.015}
\providecommand{\NFLGlobalDyadicP}{\ensuremath{0.736}}
\providecommand{\NFLLOSODMax}{0.025}
\providecommand{\NFLLOSODMin}{0.008}

\providecommand{\NFLLeaveSeasonRows}{
2018 & 1,594 & 0.017 & -0.7\% \\
2019 & 1,593 & 0.013 & -0.3\% \\
2020 & 1,593 & 0.014 & -1.0\% \\
2021 & 1,577 & 0.013 & -0.3\% \\
2022 & 1,579 & 0.008 & -0.8\% \\
2023 & 1,576 & 0.024 & -0.4\% \\
2024 & 1,576 & 0.025 & 0.0\% \\
}
\providecommand{\NFLMedianUpdatesPerGame}{175}

\providecommand{\NFLMissingFeedGames}{0}

\providecommand{\NFLPatchedGames}{4}
\providecommand{\NFLPublishedUpdates}{325,922}
\providecommand{\NFLRawGlobalD}{0.015}
\providecommand{\NFLRawGlobalP}{\ensuremath{0.735}}
\providecommand{\NFLRoundingLowerGlobalD}{0.013}

\providecommand{\NFLRoundingLowerTailNinetyFiveRate}{4.5\%}
\providecommand{\NFLRoundingUpperGlobalD}{0.016}

\providecommand{\NFLRoundingUpperTailNinetyFiveRate}{4.5\%}
\providecommand{\NFLScheduledGames}{1,856}
\providecommand{\NFLTailNinetyCILower}{-3.5\%}
\providecommand{\NFLTailNinetyCIUpper}{0.7\%}
\providecommand{\NFLTailNinetyExcess}{-1.5\%}
\providecommand{\NFLTailNinetyFiveAtomRate}{0.00\%}
\providecommand{\NFLTailNinetyFiveCILower}{-2.0\%}
\providecommand{\NFLTailNinetyFiveCIUpper}{1.1\%}
\providecommand{\NFLTailNinetyFiveExcess}{-0.5\%}
\providecommand{\NFLTailNinetyFiveRate}{4.5\%}
\providecommand{\NFLTailNinetyFiveStrictRate}{4.5\%}
\providecommand{\NFLTailNinetyNineCILower}{-0.3\%}
\providecommand{\NFLTailNinetyNineCIUpper}{1.5\%}
\providecommand{\NFLTailNinetyNineExcess}{0.5\%}
\providecommand{\NFLTailNinetyNineRate}{1.5\%}
\providecommand{\NFLTailNinetyRate}{8.5\%}

\providecommand{\NFLUpdateQOne}{167}
\providecommand{\NFLUpdateQThree}{185}
\providecommand{\SimulationBaseGrid}{1,000}

\providecommand{\ValidationInnerReplicates}{399}
\providecommand{\ValidationNBARhoFiftyDyadic}{4.4\%}
\providecommand{\ValidationNBARhoTwentyFiveDyadic}{3.2\%}
\providecommand{\ValidationNBARhoZeroDyadic}{0.0\%}
\providecommand{\ValidationNFLRhoFiftyDyadic}{3.2\%}
\providecommand{\ValidationNFLRhoTwentyFiveDyadic}{0.8\%}
\providecommand{\ValidationNFLRhoZeroDyadic}{0.0\%}
\providecommand{\ValidationOuterSamples}{250}
\theoremstyle{plain}
\newtheorem{theorem}{Theorem}

\newtheorem{proposition}[theorem]{Proposition}
\newtheorem{corollary}[theorem]{Corollary}
\theoremstyle{definition}
\newtheorem{definition}[theorem]{Definition}
\newtheorem{remark}[theorem]{Remark}
\hypersetup{pdftitle={The Blown Lead Paradox: A Pathwise Calibration Benchmark for Win Probability Forecasts},pdfauthor={Jonathan Pipping-Gamon and Abraham J. Wyner},pdfkeywords={Forecast calibration, martingales, pigeonhole bootstrap, sports analytics, win probability}}
\title{The Blown Lead Paradox\\[0.25em]\large A Pathwise Calibration Benchmark for Win Probability Forecasts}
\author{Jonathan Pipping-Gam\'on \and Abraham J. Wyner}
\date{August 25, 2026}
\begin{document}
\maketitle
\begin{abstract}
Live win probabilities have given sports collapses a common numerical language: the highest win probability attained by the team that eventually lost. Yet this familiar number is a selected pathwise extreme. The losing path is identified by the terminal outcome and then searched for its most favorable point. Under ideal sequential calibration, we derive an exact continuous-path benchmark for this statistic, a conservative bound for discretely reported paths, and a probability-integral-transform diagnostic for collections of games. Unlike fixed-time calibration summaries and proper scores, the diagnostic asks whether a published feed produces severe losing paths as often as a calibrated sequential model should. We apply the benchmark separately to public regular-season NFL and NBA feeds from 2018--2024, using a season-stratified dyadic bootstrap to account for recurring teams. We detect no global departure in the NFL and a systematic excess of extreme losing-team peaks in the NBA\@. In the upper 5\% benchmark tail, the NBA excess is \NBATailNinetyFiveExcess{} (95\% interval \NBATailNinetyFiveCILower{} to \NBATailNinetyFiveCIUpper{}); first crossings of a published win probability of $0.95$ show the same positive gap between observed and implied loss rates.
\end{abstract}
{\small\noindent\textbf{Keywords:} \mbox{forecast calibration}; martingales; \mbox{pigeonhole bootstrap}; \mbox{sports analytics}; \mbox{win probability}.\par}

\section{Introduction}

On \ChicagoDate, with \ChicagoClock{} remaining in the third quarter, the Chicago Bulls led the Los Angeles Lakers \ChicagoPeakHomeScore--\ChicagoPeakAwayScore{}.
ESPN's win probability feed put Chicago's chance of winning at \ChicagoMaxWP{}; the Bulls went on to lose \ChicagoFinalHomeScore--\ChicagoFinalAwayScore{} \citep{ESPN2019BullsLakers}.
After the final whistle, that one number became a natural summary of the collapse.
This is the appeal of win probability: a basketball lead of \ChicagoPeakLead{} points and a two-score football lead are not directly comparable on the scoreboard, but both can be expressed on the same probability scale.
Modern accounts of blown leads therefore often report the highest win probability reached by the team that eventually lost.

That convenience hides a conditioning problem.
When the \ChicagoMaxWP{} forecast was issued, its complement described Chicago's conditional chance of losing under the model.
After the game, however, Chicago was selected precisely because it lost, and its path was then searched for its most favorable point.
The complement of the selected maximum is not the frequency with which calibrated losing paths should reach a maximum that high.
Answering the retrospective question requires a different benchmark: among calibrated paths that end in defeat, how large should the losing team's maximum be?

For a two-team game, we study the largest win probability attained by whichever team eventually loses.
The terminal outcome selects one team's path, and the reported sequence supplies its observed peak.
This quantity is easy to read from a game graphic but less easy to judge, because its comparison class consists of losing paths rather than all forecasts made at a fixed probability.
If an eventual loser once reached $95\%$, the relevant question is not simply whether $5\%$ events occur.
It is how often a calibrated sequential model produces losing paths with peaks at or above $95\%$.

Let $Y\in\{0,1\}$ indicate whether Team~A wins.
Our reference model is ideal sequential calibration:
\[
p_k=\P(Y=1\mid \mathcal F_k),\qquad k=0,1,\dots,N,
\]
under which the forecast is a bounded martingale ending at $p_N=Y$.
This ideal is stronger than marginal calibration within probability bins: every probability must equal the conditional chance of winning given the information then available, so the sequence is dynamically coherent.
That martingale structure yields an exact benchmark for continuously observed paths and a conservative one for discretely reported paths.
It also turns each game into a common benchmark score that can be aggregated across a league.

The resulting diagnostic adds a pathwise question to ordinary forecast evaluation.
Reliability diagrams ask whether events forecast near $p$ occur about $p$ of the time, while Brier and log scores summarize predictive performance across individual forecasts \citep{Brier1950,Murphy1973,DeGrootFienberg1983,Dawid1982,GneitingRaftery2007}.
These tools evaluate individual forecasts; the blown-lead statistic evaluates a maximum chosen after the losing team is known.
A feed can look broadly reliable update by update and still produce too many extreme losing-team maxima, or the reverse.

That distinction connects two strands of earlier work.
Sports statisticians have modeled in-game win probability through score-evolution models \citep{Stern1994} and play-level machine-learning methods \citep{LockNettleton2014,YurkoVenturaHorowitz2019}.
Most directly, \citet{YehRiceDubin2022} developed calibration surfaces and comparative tests for continuously updated NBA forecasts and found ESPN's published feed broadly calibrated at the update level.
Their result makes the pathwise question sharper: can a feed whose individual updates look reasonable nevertheless produce too many extreme maxima among eventual losers?
The sequential-forecasting literature supplies the martingale and prequential language for asking that question \citep{Dawid1984,FosterVohra1998}, while recent work provides broader tests of sequential calibration \citep[e.g.,][]{Howard2020,RamdasGrunwaldVovkShafer2023,ArnoldHenziZiegel2023}.

The remaining obstacle is the reverse conditioning built into a blown lead.
Doob- and Ville-type inequalities control unconditional events such as $\{\sup_t p_t\ge x\}$, but the sports question selects paths that end in failure.
Exact identities for martingale suprema are known in related continuous-time settings \citep{NikeghbaliYor2006}.
Specializing that logic to bounded martingales with binary terminal values gives a closed-form law for the losing team's maximum, along with corrections for discrete reporting.

The empirical applications use public regular-season NFL and NBA win probability feeds from 2018--2024.
To account for recurring teams and matchups, we use a season-stratified dyadic pigeonhole bootstrap across \NFLGames{} NFL games and \NBAGames{} NBA games.
The NFL sample shows no detected global departure ($p=\NFLGlobalDyadicP$), whereas the NBA sample rejects the upper-tail benchmark ($p\NBAGlobalDyadicP$).
At $U\ge0.95$, \NBATailNinetyFiveRate{} of NBA games fall in a benchmark tail containing $5\%$ under the continuous null; the estimated excess is \NBATailNinetyFiveExcess{}, with a 95\% dyadic-bootstrap interval from \NBATailNinetyFiveCILower{} to \NBATailNinetyFiveCIUpper{}.

\section{The peak-on-loss benchmark} \label{sec:theory}

The conditioning problem becomes tractable once the forecast path is written as a martingale.
Let $Y\in\{0,1\}$ be revealed at time $N$, and let $p_k=\E[Y\mid\mathcal F_k]$ be the ideal sequential forecast, with $p_0\in(0,1)$ and $p_N=Y$.
For the continuous-time counterpart, let $p_t=\E[Y\mid\mathcal F_t]$ for $t\in[0,1]$, with the same initial value and $p_1=Y$.
Write
\[
M_N := \max_{0\le k\le N} p_k,
\qquad
M := \sup_{t\in[0,1]} p_t
\]
for the discrete- and continuous-time path maxima.

\begin{definition}[First-passage time]
For $x\in(0,1]$, define
\[
\tau_x := \inf\{k\in\{0,1,\dots,N\}: p_k\ge x\},
\]
with $\inf\varnothing := N+1$.
\end{definition}

The first object we need is the \emph{peak-on-loss} law: the distribution of a team's path maximum among games that team loses.

\begin{theorem}[Peak-on-loss benchmark] \label{thm:continuous-conditional}
Under sequential calibration and continuous sample paths,
\[
\P(M\ge x \mid Y=0)=\frac{p_0}{1-p_0}\cdot\frac{1-x}{x},
\qquad x\in[p_0,1).
\]
The displayed tail is continuous in $x$ below one, so the conditional maximum has no atoms there.
Equivalently, for $x\in[p_0,1)$,
\[
F_{M\mid Y=0}(x)=1-\frac{p_0}{1-p_0}\cdot\frac{1-x}{x},
\]
with $F_{M\mid Y=0}(x)=0$ for $x<p_0$ and $F_{M\mid Y=0}(1)=1$.
\end{theorem}

\begin{theorem}[Discrete-time peak-on-loss bound] \label{thm:discrete-conditional}
In discrete time,
\[
\P(M_N<x\mid Y=0)\ \ge\ 1-\frac{p_0}{1-p_0}\cdot\frac{1-x}{x},
\qquad x\in[p_0,1),
\]
and therefore the same lower bound holds for the ordinary CDF $F_{M_N\mid Y=0}(x)=\P(M_N\le x\mid Y=0)$.
The strict-threshold bound is exact when the path neither overshoots $x$ at first passage nor first attains $x$ at the terminal reveal. Equality for the ordinary CDF additionally requires \(\P(M_N=x\mid Y=0)=0\).
\end{theorem}

\begin{remark}[Discrete-time correction identity] \label{rem:discrete-correction}
For $x\in(p_0,1)$,
\[
\P(M_N\ge x\mid Y=0)
=\frac{p_0}{1-p_0}\cdot\frac{1-x}{x}\;-\;C_1(x)\;-\;C_2(x),
\]
where
\[
C_1(x):=\frac{1}{x(1-p_0)}\,\E\!\big[(p_{\tau_x}-x)\,\mathbf 1\{\tau_x< N\}\big],
\qquad
C_2(x):=\frac{1-x}{x(1-p_0)}\,\P(\tau_x=N).
\]
Here $C_1$ measures preterminal first-passage overshoots, while $C_2$ measures paths that first attain $x$ only when the outcome is revealed.
Both terms are nonnegative and quantify how discrete first passage lowers the tail relative to the continuous-path benchmark.
\end{remark}

Proofs and discrete-time corrections are given in Appendices~\ref{app:discrete-proofs}--\ref{app:two-player-derivation}.

\section{A game-level benchmark score} \label{sec:applications}

Theorem~\ref{thm:continuous-conditional} gives the law of a single team's path maximum conditional on that team losing.
The sports graphic poses the question in reverse: begin with a game, identify the loser after the outcome is known, and then read off that team's best moment.
Combining the two possible losing paths turns the conditional theorem into a benchmark for a complete game.

Consider a two-team contest and encode the outcome as $Y\in\{0,1\}$, where $Y=1$ means Team~A wins and $Y=0$ means Team~B wins.
Let
\[
p_t=\P(Y=1\mid\mathcal F_t)
\qquad\text{and}\qquad
q_t=1-p_t=\P(Y=0\mid\mathcal F_t)
\]
denote the win probability martingales for Teams~A and B.
The \emph{loser's peak win probability} is the maximum win probability attained by the team that ultimately loses:
\[
M_\lambda :=
\sup_{t\in[0,1]}\Big(p_t\,\mathbf 1\{Y=0\}+q_t\,\mathbf 1\{Y=1\}\Big).
\]
We reserve $M_\lambda$ for this ideal continuous-path maximum; lowercase $m_i$ will denote the peak observed on game $i$'s published path.
On $\{Y=0\}$, Team~A loses and $M_\lambda=\sup_t p_t$; on $\{Y=1\}$, Team~B loses and $M_\lambda=\sup_t q_t$.
We therefore apply Theorem~\ref{thm:continuous-conditional} to each team's losing path and mix over the two outcomes; Appendix~\ref{app:two-player-derivation} gives the full derivation.

\begin{figure}[!tbp]
\centering
\includegraphics[width=0.9\linewidth]{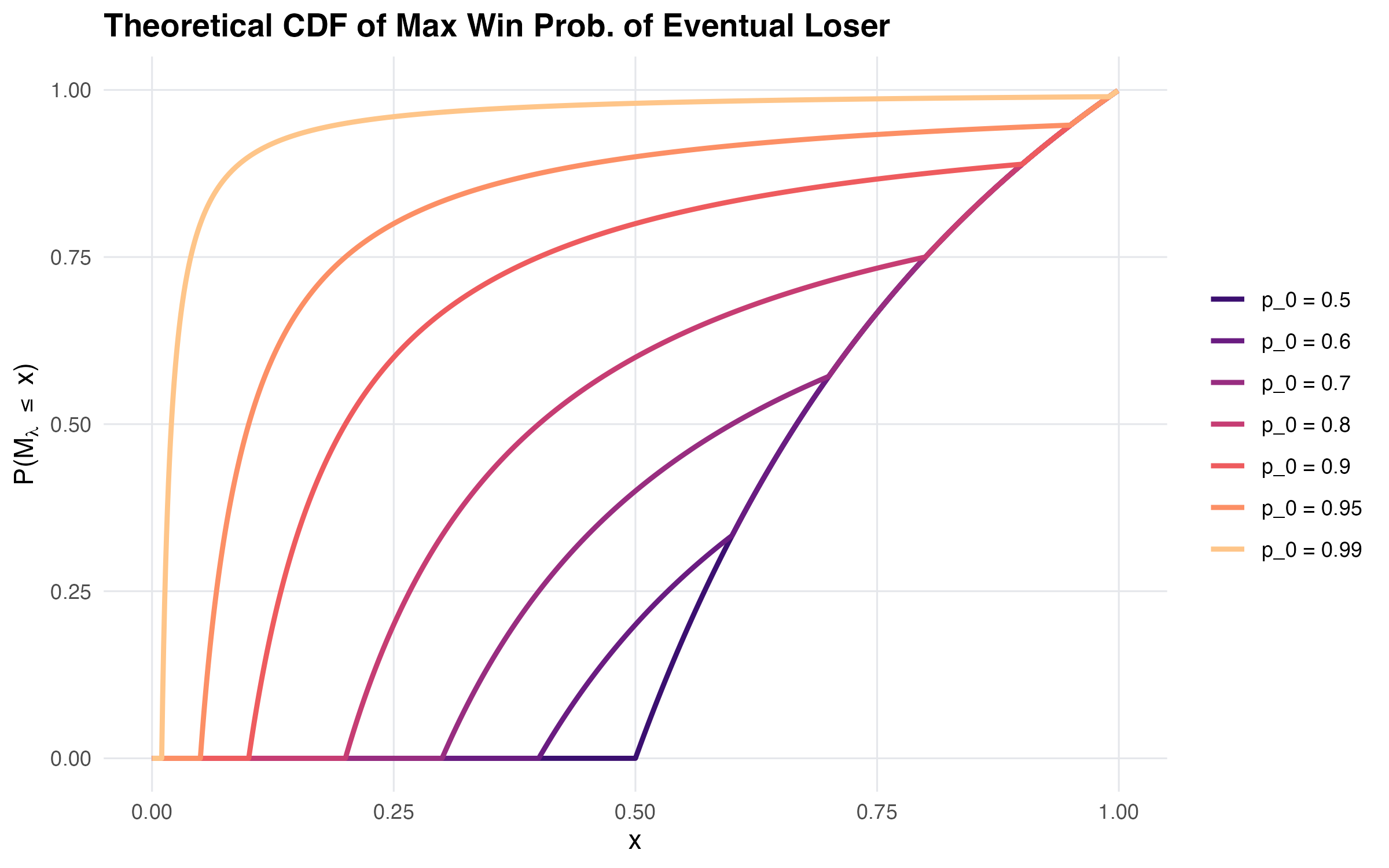}
\caption{Benchmark CDF $F_{M_\lambda}(x)$ for different pre-game favorite probabilities $p_0$.}
\label{fig:sports-cdf}
\end{figure}

Assume without loss of generality that Team~A is the pre-game favorite, so $p_0\ge 1/2$.
Under continuous sample paths, the benchmark CDF is
\[
F_{M_\lambda}(x)=
\begin{cases}
0 & x<1-p_0,\\[0.3em]
1-\dfrac{1-p_0}{x} & 1-p_0\le x<p_0,\\[0.6em]
2-\dfrac{1}{x} & p_0\le x<1,\\[0.3em]
1 & x=1.
\end{cases}
\]
The three regimes reflect which losing path can fall below $x$.
Below $1-p_0$, neither can; between $1-p_0$ and $p_0$, only the underdog's losing path can; above $p_0$, both can.
In this final region the tail no longer depends on the starting probability:
\[
\P(M_\lambda\ge x)=\frac{1}{x}-1,\qquad x\in[p_0,1).
\]
In particular, when $p_0\le 0.9$, $\P(M_\lambda\ge 0.9)=1/9\approx 11\%$.
When $p_0\ne 1/2$, the intermediate region $[1-p_0,p_0)$ produces a kink (Figure~\ref{fig:sports-cdf}).

\begin{figure}[!tbp]
\centering
\includegraphics[width=0.92\linewidth]{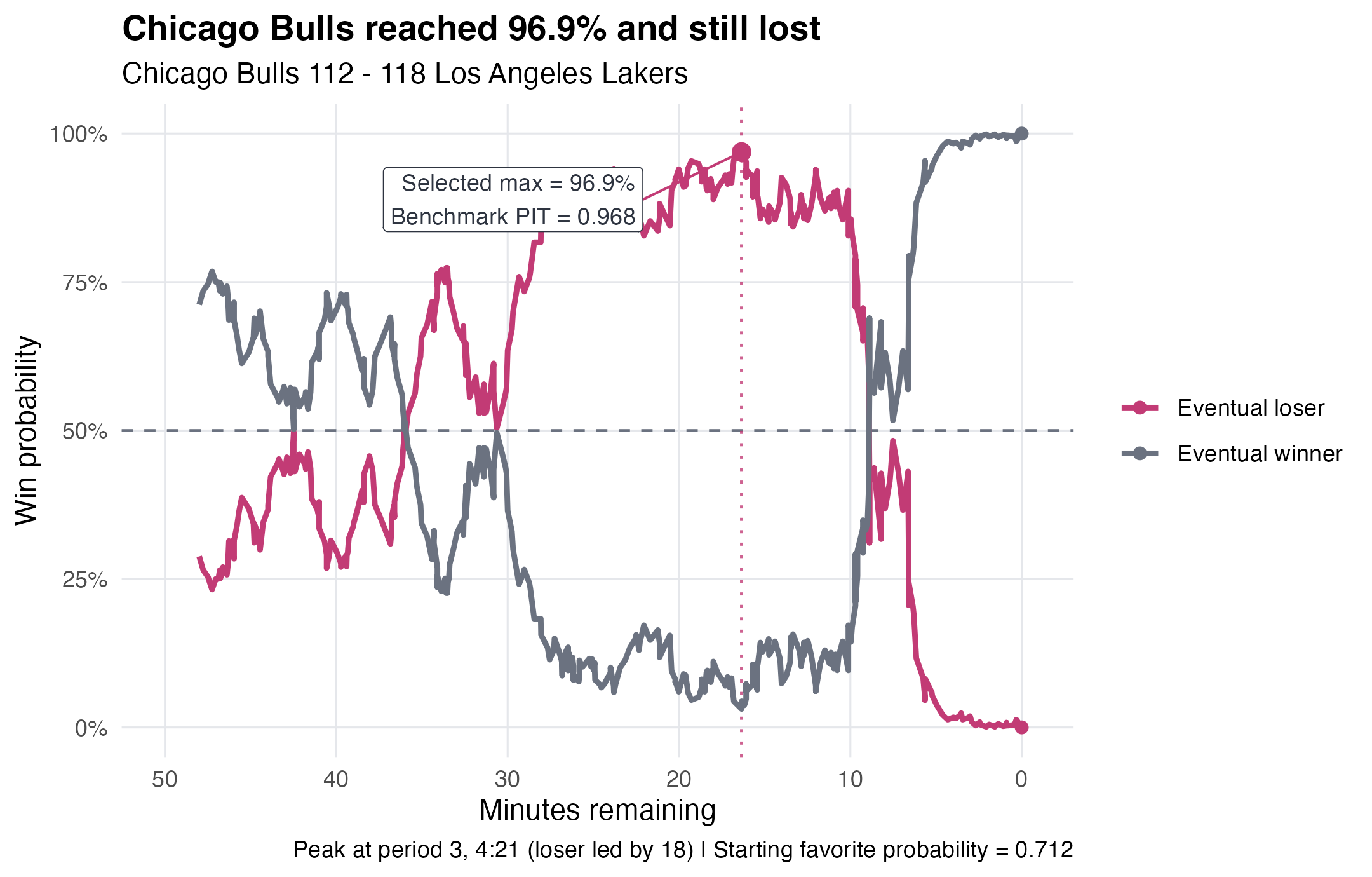}
\caption{Chicago's published win probability paths. The eventual loser's observed peak is \ChicagoMaxWP{} with \ChicagoMinutesRemaining{} minutes remaining; the game-specific benchmark maps that peak and $p_0=\ChicagoStartingWP$ to $U=\ChicagoPIT$.}
\label{fig:nba-extreme-case}
\end{figure}

This distribution gives the retrospective statement ``the eventual loser reached $90\%$'' its proper reference point.
Such an event need not be rare under a calibrated live model: when the pre-game favorite begins at or below $90\%$, the benchmark probability is about $11\%$.
The comeback may still be surprising when it occurs.
What changes is the comparison set: the selected maximum must be judged against other game paths that also end in defeat, not against forecasts observed at a fixed $90\%$.
For the Chicago game, $p_0=\ChicagoStartingWP$, the observed published-path maximum is $m=\ChicagoMaxWP$, and the benchmark score is $U=\ChicagoPIT$, with upper-tail probability \ChicagoTailProbability{} (Figure~\ref{fig:nba-extreme-case}).

Chicago is unusual even on this scale, but one game cannot diagnose a feed.
The league-level question is whether games at least this unusual recur at the rate the benchmark allows.
Appendix~\ref{sec:supp-benchmarks} gives benchmark values for other common thresholds.

\section{A league-level diagnostic} \label{sec:diagnostics}

Once a single game can be scored, the next question is how to compare many such games drawn from different starting conditions.
The benchmark CDF puts their losing-team peaks on a common scale; the distribution of those scores then reveals whether a feed produces too many severe collapses.

\subsection{Game-level PIT scores}

For game $i$, let $p_{0,i}$ be the first reported favorite probability and let $m_i$ be the observed maximum win probability assigned to the team that eventually lost.
The game-specific benchmark score is
\[
U_i := F_{M_\lambda}(m_i; p_{0,i}).
\]
We call $U_i$ the probability-integral-transform (PIT) score associated with the game-specific benchmark.
Thus $M_\lambda$ is the ideal continuous losing-team peak, $m_i$ is the observed published-path peak, $U_i$ scores one game, and $D_n^{\mathrm{upper}}$ below aggregates those scores across a league.
Large values of $U_i$ correspond to unusually severe collapses under the benchmark.
For example, $U_i=0.97$ means that the losing-team peak exceeds 97\% of benchmark realizations with the same starting favorite probability.

Under the continuous-path law, $U_i$ is $\mathrm{Unif}(0,1)$.
For a discretely reported feed, missed peaks move scores away from the upper tail, so an excess of large $U_i$ remains the direction of interest.

\subsection{Dependence-robust league inference}

At the league level, the question is whether the empirical distribution of the $U_i$ places too much mass near one.
Under the continuous-path benchmark, the scores are uniform even when starting probabilities vary from game to game.
Discrete reporting moves scores downward, so for $\alpha\in(0,1)$ the operative null is
\[
\P(U_i\ge 1-\alpha)\le \alpha.
\]
The global statistic follows that direction:
\[
D_n^{\mathrm{upper}}
  :=\sup_{0\le t\le1}\{t-\hat F_U(t-)\}.
\]
We call the curve $t-\hat F_U(t-)$ the centered PIT signature; $D_n^{\mathrm{upper}}$ is its maximum.
The left limit preserves the weak upper-tail convention when published probabilities create ties; computationally, we maximize $u-\hat F_U(u-)$ over the distinct observed scores.
A large value means that the empirical CDF accumulates too slowly because too much mass lies near one.

Recurring teams and repeat matchups create shared-team dependence within seasons.
We account for this shared-team dependence with a season-stratified version of the pigeonhole bootstrap \citep{Owen2007,DaveziesDHAultfoeuilleGuyonvarch2021}.
Within season $s$, let $W_{sa}$ be the multiplicity obtained by sampling that season's teams multinomially with replacement.
A game between teams $a$ and $b$ receives weight $W_{sa}W_{sb}$.
Thus repeat meetings share an endpoint weight, while normalization makes the total weight in each season equal its observed number of games.
The corresponding asymptotic argument assumes that each season is a jointly exchangeable, dissociated dyadic array: games may depend when they share a team but are independent when their endpoint sets are disjoint.
The multiple-observation extension of \citet{DaveziesDHAultfoeuilleGuyonvarch2021} allows both zero games and repeated games in a matchup cell.

For each of $B=\DyadicBootstrapReplicates$ draws, we recompute the weighted empirical CDF and form the centered statistic
\[
D_n^*=\sup_t\{\hat F_U(t)-\hat F_U^*(t)\}.
\]
The reported global $p$-value is
\[
\frac{1+\sum_{b=1}^B \mathbf 1\{D_{n,b}^*\ge D_n^{\mathrm{upper}}\}}{B+1}.
\]
Recentering measures how far a resampled CDF can fall below the observed CDF without carrying the observed departure into every replicate.
We pair this global test with an interpretable effect size,
\[
\Delta_{.95}:=\widehat{\P}_n(U\ge0.95)-0.05.
\]
Its uncertainty is summarized by a two-sided percentile interval from the same dyadic draws.
Under these conditions, the recentered pigeonhole bootstrap consistently estimates the CDF empirical process.
Together with $F_U(t-)\ge t$, this yields an asymptotically conservative upper-tail test.
Because this result is asymptotic, we assess finite-schedule behavior using simulations on the observed schedules.
Appendix~\ref{sec:supp-inference} specializes the exchangeable-array bootstrap result to the normalized game-level CDF; Appendices~\ref{sec:supp-results}--\ref{sec:supp-validation} report the dependence sensitivities and finite-schedule null simulation.

\section{Application to NFL and NBA win probability feeds} \label{sec:empirical}

We apply the game-level score separately to ESPN's public NFL and NBA win probability paths \citep{ESPN2025}.
The NFL data were assembled with \texttt{espnscrapeR} \citep{espnscrapeR2025}; the NBA data were assembled with \texttt{hoopR} \citep{hoopR2023}.
Both samples contain regular-season franchise games only, although the season labels follow league convention.
An NFL season is labeled by the year in which it begins, so games played the following January retain the preceding year's label; an NBA season is labeled by the year in which it ends.
Because ESPN's NBA season-type field also includes All-Star exhibitions, we additionally require both participants to be NBA franchises.
That filter removes \NBAExcludedExhibitionEvents{} All-Star or Rising Stars events, \NBAExcludedExhibitionFeeds{} of which had usable probability paths.
From the 2018--2024 season inventories, we retain completed games, exclude NFL ties, and require a usable published probability path.
Table~\ref{tab:data-coverage} shows the resulting coverage.

\begin{table}[!tbp]
\centering
\small
\caption{Regular-season feed construction. Eligible games are completed, non-tied games; analyzed games also have a usable public win probability path.}
\label{tab:data-coverage}
\begin{tabular}{lrrrrr}
\hline
League & Scheduled & Completed & Eligible & Analyzed & Missing feed \\
\hline
NFL & \NFLScheduledGames & \NFLCompletedGames & \NFLEligibleGames & \NFLGames & \NFLMissingFeedGames \\
NBA & \NBAScheduledGames & \NBACompletedGames & \NBAEligibleGames & \NBAGames & \NBAMissingFeedGames \\
\hline
\end{tabular}
\end{table}

Our estimand is ESPN's published feed, encompassing both forecast generation and the conversion of internal output into displayed probabilities.
For each game we retain the first published probability, the complete sequence of published updates, the final outcome, and the first attainment of the eventual loser's peak.
Overtime updates remain part of the collapse statistic because a collapse can occur in overtime.

The feeds are published at different resolutions.
The NFL sample contains \NFLPublishedUpdates{} updates, with a median of \NFLMedianUpdatesPerGame{} per game (interquartile range \NFLUpdateQOne--\NFLUpdateQThree); the NBA sample contains \NBAPublishedUpdates{}, with a median of \NBAMedianUpdatesPerGame{} (interquartile range \NBAUpdateQOne--\NBAUpdateQThree).
We preserve those resolutions because thinning a path can hide its maximum and change the estimand.
The leagues serve as parallel applications of the same benchmark, with each interpreted in the context of its own schedule, game dynamics, and publication resolution.

\subsection{Fixed-clock calibration}

Before asking how high the losing path ever rose, consider the more familiar question studied by \citet{YehRiceDubin2022}: at comparable points in a game, do published probabilities agree with observed win rates?
We adapt their calibration-surface idea to a common regulation clock.
Each game's home-team forecast is evaluated at $t=0,.05,\ldots,.95$ of regulation, using linear interpolation between published updates, with piecewise-constant interpolation as a sensitivity analysis.
At each time, ten adaptive probability bins are formed from empirical quantiles without splitting tied forecasts.
Every retained game therefore contributes once at each time rather than in proportion to its number of published updates.
Overtime is excluded from this common-clock display.
The complete fixed-clock samples contain \NFLFixedClockGames{} NFL games and \NBAFixedClockGames{} NBA games; \NBAFixedClockExcluded{} NBA paths end before the full grid and are excluded from this comparison only.

\begin{figure}[!tbp]
\centering
\includegraphics[width=0.95\linewidth]{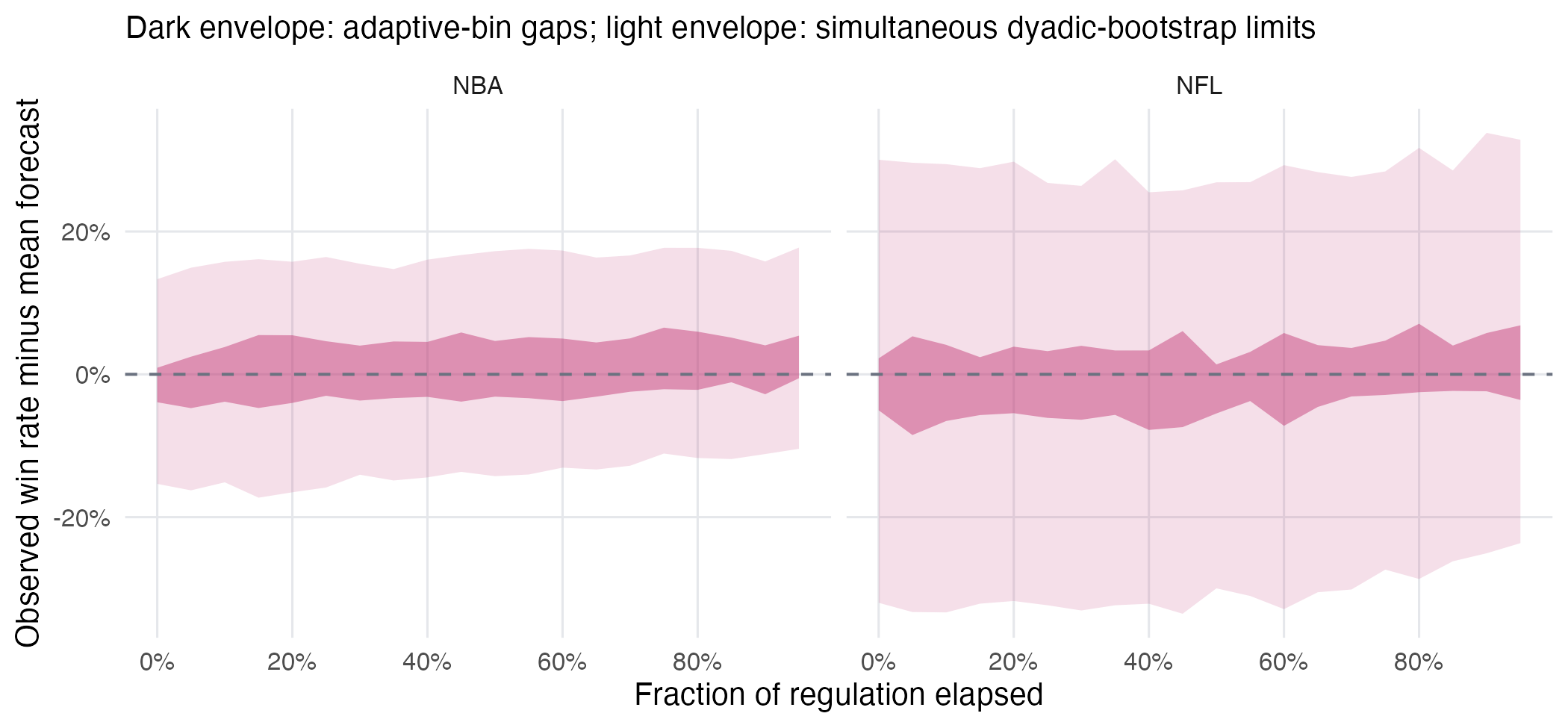}
\caption{Fixed-clock calibration envelopes using linear interpolation. Each game contributes once at each fraction of regulation. Dark regions span the ten tie-preserving adaptive-bin gaps; light regions include the simultaneous season-stratified dyadic-bootstrap limits. Full time--probability surfaces and the LOCF sensitivity appear in the supplement.}
\label{fig:fixed-clock-envelope}
\end{figure}

Figure~\ref{fig:fixed-clock-envelope} compresses the resulting calibration surfaces.
At each time, the dark region spans the observed-minus-forecast gaps across bins, and the lighter region spans their simultaneous dyadic-bootstrap limits.
In the full time--probability surfaces, the cells whose simultaneous intervals exclude zero are concentrated late in regulation and near published probabilities of zero or one in both applications (Appendix Figure~\ref{fig:supp-fixed-clock}).
The two diagnostics answer different questions: fixed-clock calibration evaluates forecasts at prespecified times, whereas the losing-path diagnostic selects a maximum after the loser is known.

\subsection{The losing-path diagnostic}

\begin{figure}[!tbp]
\centering
\begin{minipage}[t]{0.8\linewidth}
  \centering
  \includegraphics[width=\linewidth]{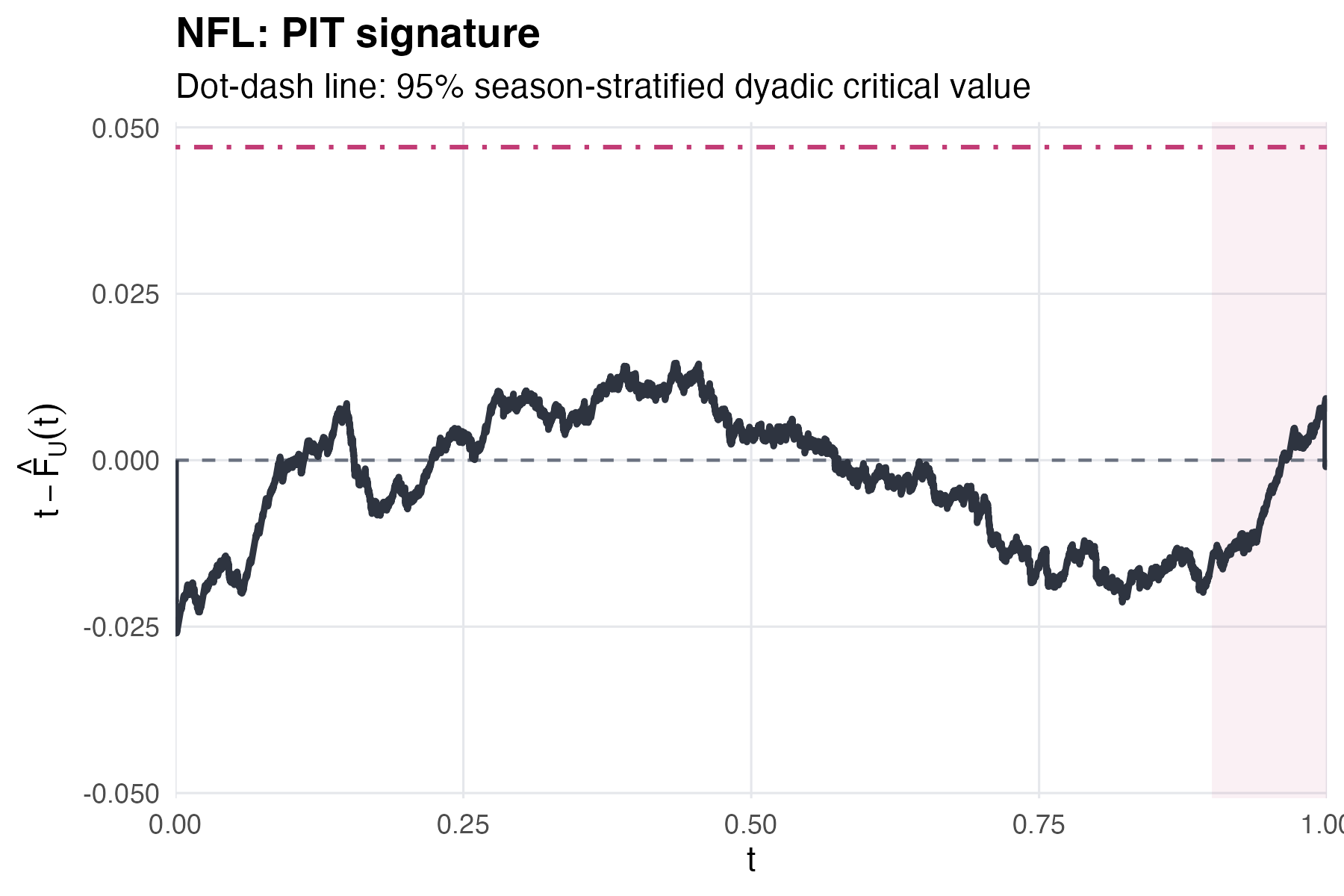}
  \smallskip\par\small (a) NFL
\end{minipage}\\[0.5em]
\begin{minipage}[t]{0.8\linewidth}
  \centering
  \includegraphics[width=\linewidth]{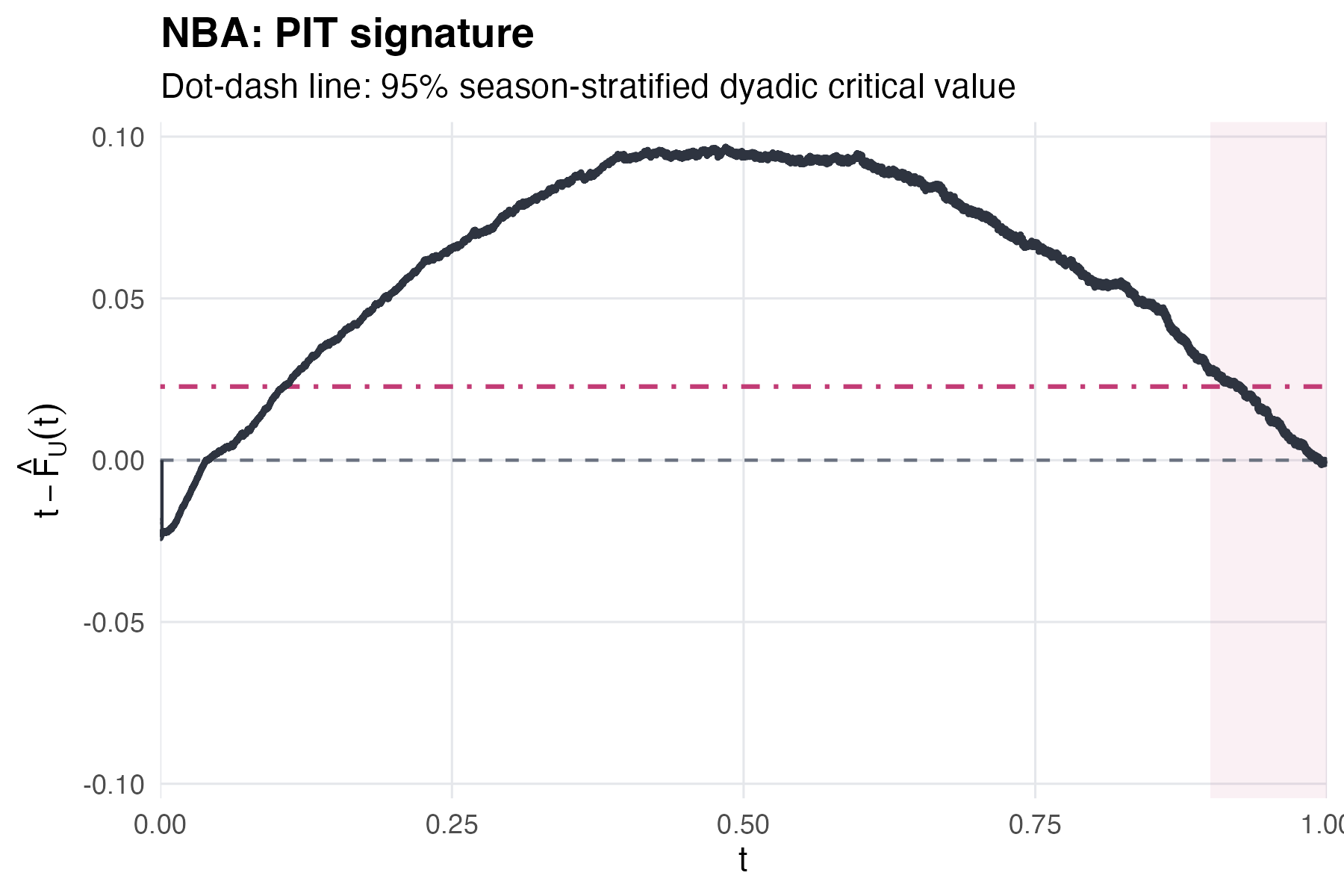}
  \smallskip\par\small (b) NBA
\end{minipage}
\caption{Centered PIT signatures $t-\hat F_U(t-)$ and the 95\% season-stratified dyadic-bootstrap critical value (dot-dash). The NFL application has no detected global departure; the NBA application exceeds its reference.}
\label{fig:pit-combined}
\end{figure}

For each game, the losing-path diagnostic maps the eventual loser's peak $m_i$ through the game-specific benchmark CDF before aggregation.
Figure~\ref{fig:pit-combined} shows the resulting curves $t-\hat F_U(t-)$.
The dot-dash line in each panel marks the 95th percentile of the centered dyadic-bootstrap statistic used for formal inference.

Table~\ref{tab:primary-inference} reports the global test and the $0.95$ tail magnitude for each application.

\begin{table}[!tbp]
\centering
\small
\caption{Primary inference for each application. The interval is a two-sided percentile interval from the same season-stratified dyadic draws used for the global test.}
\label{tab:primary-inference}
\begin{tabular}{lccccc}
\hline
League & Games & $D_n^{\mathrm{upper}}$ & Dyadic $p$ & $\widehat{\P}(U\ge0.95)$ & $\Delta_{.95}$ (95\% CI) \\
\hline
NFL & \NFLGames & \NFLGlobalD & \NFLGlobalDyadicP & \NFLTailNinetyFiveRate & \NFLTailNinetyFiveExcess{} [\NFLTailNinetyFiveCILower, \NFLTailNinetyFiveCIUpper] \\
NBA & \NBAGames & \NBAGlobalD & \NBAGlobalDyadicP & \NBATailNinetyFiveRate & \NBATailNinetyFiveExcess{} [\NBATailNinetyFiveCILower, \NBATailNinetyFiveCIUpper] \\
\hline
\end{tabular}
\end{table}

The two applications diverge.
For the NFL, $D_n^{\mathrm{upper}}=\NFLGlobalD$ and the dyadic test gives $p=\NFLGlobalDyadicP$.
This conservative test yields no detected departure, leaving exact calibration unresolved.
For the NBA, $D_n^{\mathrm{upper}}=\NBAGlobalD$ and the upper-tail null is rejected ($p\NBAGlobalDyadicP$).
We use the global test to detect departure over the full score distribution and the $0.95$ tail to express its magnitude on a familiar scale.
In the NBA, \NBATailNinetyFiveRate{} of games occupy a benchmark tail of size $5\%$, an excess of \NBATailNinetyFiveExcess{} (95\% interval \NBATailNinetyFiveCILower{} to \NBATailNinetyFiveCIUpper{}).

\FloatBarrier
\subsection{First crossings of 0.95}

The event $U\ge0.95$ refers to the upper 5\% of the \emph{game-level collapse benchmark}; it is not the event that a published win probability crosses $0.95$.
The latter offers a more immediate view of where the discrepancy enters.
For a team-side path, define
\[
\tau_{.95}:=\inf\{k:p_k\ge0.95,\ \text{game time remaining}>0\}.
\]
If no preterminal crossing occurs, set $\tau_{.95}=N$.
Then $\tau_{.95}$ is a bounded stopping time and the crossing event $\{\tau_{.95}<N\}$ is known at that time.
Sequential calibration gives
\[
\E\!\left[(p_{\tau_{.95}}-Y)
  \mathbf 1\{\tau_{.95}<N\}\right]=0.
\]
Optional stopping therefore centers the selected preterminal first-crossing residual at zero under sequential calibration.
At a crossing, $1-p_{\tau_{.95}}$ is the implied loss probability and $1-Y$ is the loss indicator, so observed minus implied loss is
\[
R=(1-Y)-(1-p_{\tau_{.95}})=p_{\tau_{.95}}-Y.
\]

The NBA contributes \NBACrossingNinetyFiveEpisodes{} team-side episodes from \NBACrossingNinetyFiveGames{} games.
The observed loss rate is \NBACrossingNinetyFiveLoss{}, compared with a mean implied loss of \NBACrossingNinetyFiveImplied{}, leaving $\bar R=\NBACrossingNinetyFiveResidual$ (95\% interval \NBACrossingNinetyFiveCILower{} to \NBACrossingNinetyFiveCIUpper{}).
For the NFL, $\bar R=\NFLCrossingNinetyFiveResidual$ (\NFLCrossingNinetyFiveCILower{} to \NFLCrossingNinetyFiveCIUpper{}).
Every episode from the same game receives the same game weight, including the uncommon case in which both teams cross.
The crossing analysis shows that the NBA gap is already visible at the first entrance into a high-probability state.

\subsection{Sensitivity to feed artifacts}

The principal artifact checks change one feature at a time while reusing the same bootstrap draws.
For the NBA, $D_n^{\mathrm{upper}}$ is \NBAGlobalD{} on corrected paths, \NBARawGlobalD{} on raw paths, and \NBAExcludePatchedGlobalD{} after removing all \NBAPatchedGames{} games in which the feed incorrectly assigned probability one to the eventual loser.
Allowing each displayed probability to lie within $\pm0.0005$ bounds $D_n^{\mathrm{upper}}$ between \NBARoundingLowerGlobalD{} and \NBARoundingUpperGlobalD{}.
The dyad-plus-calendar sensitivity gives $p\NBADyadicCalendarGlobalP$, and the franchise-linked cross-season sensitivity gives $p\NBAFranchiseLinkedGlobalP$.
Leaving out one season at a time yields NBA statistics from \NBALOSODMin{} to \NBALOSODMax{}.
The NFL continues to show no detected departure under the corresponding corrections, rounding bounds, calendar blocks, franchise linkage, and season omissions.

Coverage is complete for the NFL and at least \NBAMinimumFeedCoverage{} in every NBA season.
Appendix~\ref{sec:supp-results} collects the full specification table, including boundary and terminal conventions, the $0.90$, $0.95$, and $0.99$ tail and crossing results, and leave-one-season-out analysis.
Appendix~\ref{sec:supp-data} gives the full fixed-clock surfaces, LOCF sensitivity, and season-level feed inventories.

\section{Effect of discrete reporting} \label{sec:simulation}

The continuous benchmark is exact, but a public feed reveals only a finite set of points on the path.
Theorem~\ref{thm:discrete-conditional} says that discrete reporting can hide the true maximum and make the upper-tail diagnostic conservative.
The following simulation makes that direction visible.

For trajectory $i$, draw $p_{0,i}\sim\mathrm{Unif}(0.25,0.75)$ and $Y_i\sim\mathrm{Bernoulli}(p_{0,i})$.
With an independent standard Brownian bridge $B_i^0$, define
\[
X_i(t)=tY_i+\sigma B_i^0(t),\qquad \sigma=0.5,
\]
and, for $t<1$,
\[
p_i(t)=\operatorname{logit}^{-1}\!\left\{
\operatorname{logit}(p_{0,i})+
\frac{X_i(t)-t/2}{\sigma^2(1-t)}
\right\},
\qquad p_i(1)=Y_i.
\]
Here $p_i(t)$ is the posterior probability that the bridge ends at one given the signal observed through time $t$, so it satisfies the sequential-calibration ideal.
We simulate the process on a grid with $K_{\mathrm{base}}=\SimulationBaseGrid$ points, then retain every $m$th report before computing the eventual loser's maximum and benchmark score.

\begin{figure}[!tbp]
\centering
\includegraphics[width=0.9\linewidth]{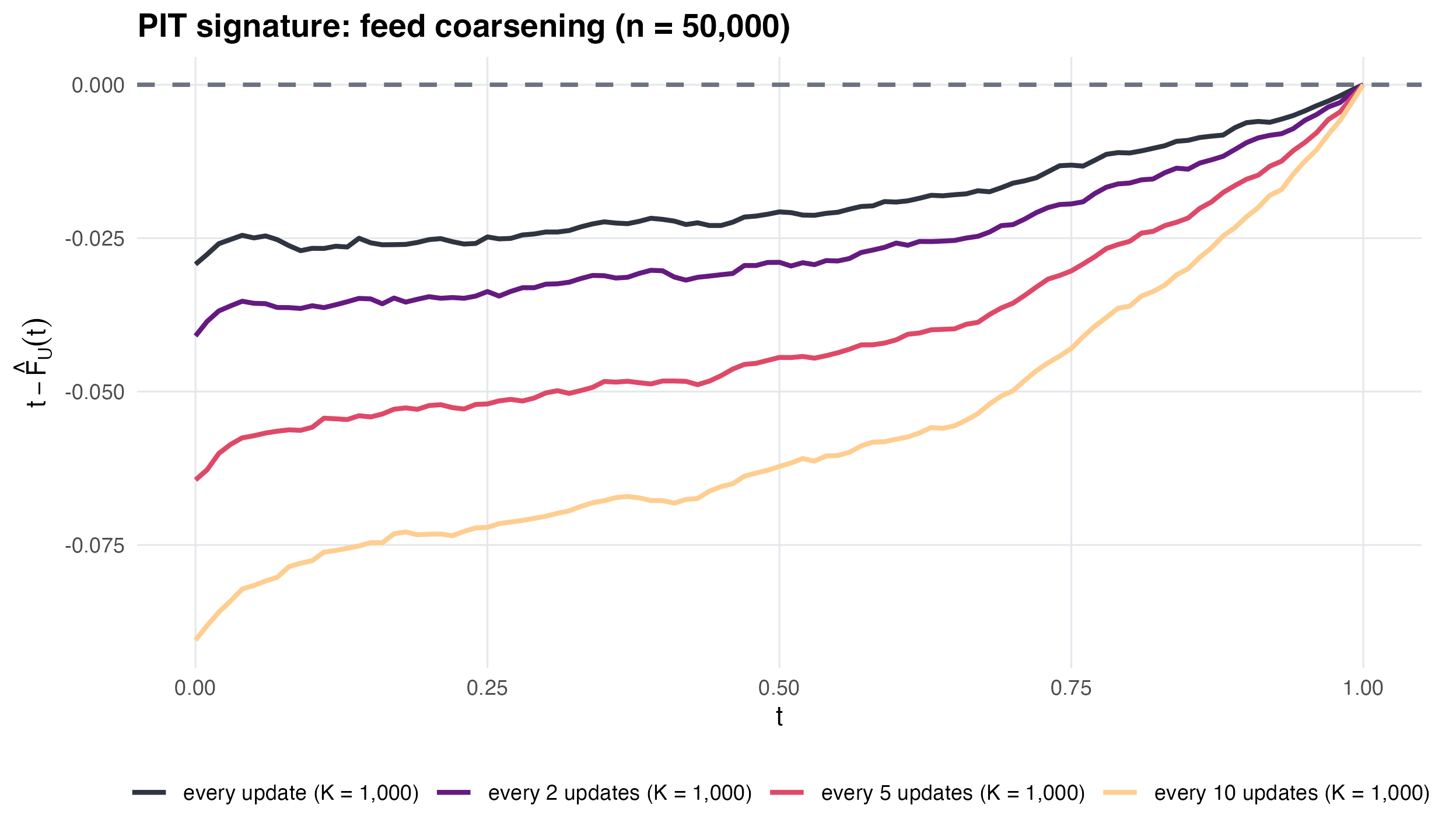}
\caption{Centered PIT signatures after observing a common calibrated process at progressively coarser resolutions ($K_{\mathrm{base}}=\SimulationBaseGrid$). Coarsening makes the upper-tail benchmark conservative.}
\label{fig:sim-null-gap}
\end{figure}

Figure~\ref{fig:sim-null-gap} shows the consequence.
As reports become less frequent, $t-\hat F_U(t-)$ moves downward.
Coarsening removes opportunities to observe a high losing-team peak and therefore works against upper-tail rejection.

\FloatBarrier
\section{Discussion} \label{sec:discussion}

A blown lead poses two questions: at $90\%$, calibration compares outcomes across similar forecasts; after the game, the selected loser's maximum is judged against the law of losing-team maxima.
That second benchmark turns a familiar sports claim into a testable statement about a published forecast sequence.
In our applications, the NFL feed shows no detected global departure, while the NBA feed contains too many extreme losing-team peaks.
The NBA departure characterizes the published feed; model specification, the effective information set, league dynamics, and the publication process can all contribute.

\subsection{Limitations and assumptions}

Our theoretical benchmark concerns the ideal process $p_k=\E[Y\mid\mathcal F_k]$.
Published win probabilities approximate this conditional expectation using an effective information set that may differ from the outcome-relevant filtration.

The exact identities additionally require continuous sample paths, which preclude first-passage overshoots and, below one, terminal-time first attainment.
For discretely reported feeds, the bounds remain valid; the explicit correction terms measure the effects of overshoots and terminal attainment.
Finally, we treat binary terminal outcomes $p_N\in\{0,1\}$; extensions to ties or other non-binary terminal structures require different arguments.

The empirical intervals describe games and matchups in the observed 2018--2024 regular seasons.
With seven season labels, season-to-season persistence is summarized descriptively.
Weak or degenerate shared-team dependence can make the pigeonhole bootstrap conservative \citep{Menzel2021}.
The finite-schedule null simulation evaluates Type I error under shared-team dependence, while franchise-linked and calendar-block sensitivities probe dependence beyond the primary dyadic model.

The Chicago game contributes one severe score, $U=\ChicagoPIT$; the recurrence of similarly extreme losing paths across \NBAGames{} NBA games is what turns a memorable collapse into evidence about a published forecasting system.

\section*{Acknowledgments}

We thank Jiaoyang Huang, Ryan Brill, and Paul Sabin for helpful discussions and feedback.

\clearpage
\appendix
\renewcommand{\thesection}{S\arabic{section}}
\renewcommand{\thesubsection}{S\arabic{section}.\arabic{subsection}}
\counterwithin{theorem}{section}
\counterwithin{table}{section}
\counterwithin{figure}{section}
\renewcommand{\thetable}{\thesection.\arabic{table}}
\renewcommand{\thefigure}{\thesection.\arabic{figure}}
\setcounter{section}{0}
\section{Proofs of Discrete-Time Results} \label{app:discrete-proofs}

Throughout, $(p_k)_{k=0}^N$ is a bounded martingale with $p_0\in(0,1)$ and terminal value $p_N=Y\in\{0,1\}$.
For $x\in(0,1]$, define the first-passage time
\[
\tau_x := \inf\{k\in\{0,1,\dots,N\}: p_k\ge x\},
\]
with $\inf\varnothing := N+1$.

\subsection{Unconditional distribution of \texorpdfstring{$M_N$}{M N}}

\begin{theorem}[Discrete-time unconditional bound] \label{thm:supp-discrete-unconditional}
Let $M_N=\max_{0\le k\le N} p_k$. For $x\in[p_0,1)$,
\[
\P(M_N<x)\;\ge\;1-\frac{p_0}{x},
\]
and hence $F_{M_N}(x)=\P(M_N\le x)\ge 1-p_0/x$.
The strict-threshold bound is an equality if $p_{\tau_x}=x$ a.s.\ on $\{\tau_x\le N\}$; equality for $F_{M_N}(x)$ additionally requires $\P(M_N=x)=0$.
For $x<p_0$, $F_{M_N}(x)=0$.
Moreover, $M_N$ has an atom at $1$ with $\P(M_N=1)=p_0$.
\end{theorem}

\begin{proof}
If $x<p_0$ then $M_N\ge p_0>x$ a.s., hence $F_{M_N}(x)=0$.

Fix $x\in[p_0,1)$. By optional stopping for the bounded martingale $(p_k)$ at $\tau_x\wedge N$,
\[
\E[p_{\tau_x\wedge N}] = p_0.
\]
Decompose $p_{\tau_x\wedge N}=p_{\tau_x}\mathbf 1\{\tau_x\le N\}+p_N\mathbf 1\{\tau_x>N\}$.
On $\{\tau_x\le N\}$ we have $p_{\tau_x}\ge x$.
Therefore,
\[
p_0
= \E[p_{\tau_x}\mathbf 1\{\tau_x\le N\}] + \E[p_N\mathbf 1\{\tau_x>N\}]
\;\ge\; x\,\P(\tau_x\le N).
\]
Since $\{\tau_x\le N\}=\{M_N\ge x\}$, this yields
\[
\P(M_N\ge x)\le \frac{p_0}{x}\qquad\Rightarrow\qquad
\P(M_N<x)\ge 1-\frac{p_0}{x}.
\]
Equality in the strict-threshold bound holds when $p_{\tau_x}=x$ a.s.\ on $\{\tau_x\le N\}$.
Because $F_{M_N}(x)=\P(M_N<x)+\P(M_N=x)$, the corresponding CDF equality also requires no atom at $x$.

Finally, $M_N=1$ iff $p_N=1$ iff $Y=1$, so $\P(M_N=1)=p_0$.
\end{proof}

\subsection{Conditional distribution of \texorpdfstring{$M_N$ given $Y=0$}{M N given Y=0}}

\begin{theorem}[Discrete-time conditional bound] \label{thm:supp-discrete-conditional}
For $x\in[p_0,1)$,
\[
\P(M_N<x\mid Y=0)
\;\ge\;
1-\Big(\frac{p_0}{1-p_0}\Big)\Big(\frac{1-x}{x}\Big),
\]
and hence the same bound holds for $F_{M_N\mid Y=0}(x)=\P(M_N\le x\mid Y=0)$.
The strict-threshold bound is an equality under the same no-overshoot condition as in Theorem~\ref{thm:supp-discrete-unconditional}; CDF equality additionally requires $\P(M_N=x\mid Y=0)=0$.
For $x<p_0$, $F_{M_N\mid Y=0}(x)=0$.
\end{theorem}

\begin{proof}
Fix $x\in[p_0,1)$. Decompose $\P(M_N\ge x)$ by $Y$:
\[
\P(M_N\ge x)=\P(M_N\ge x, Y=1)+\P(M_N\ge x, Y=0).
\]
On $\{Y=1\}$ we have $p_N=1\ge x$, hence $\{Y=1\}\subseteq\{M_N\ge x\}$ and
\begin{align*}
\P(M_N\ge x)
&=\P(Y=1)+\P(Y=0)\P(M_N\ge x\mid Y=0) \\
&=p_0+(1-p_0)\P(M_N\ge x\mid Y=0).
\end{align*}
By Theorem~\ref{thm:supp-discrete-unconditional}, $\P(M_N\ge x)\le p_0/x$, so
\[
\frac{p_0}{x}\;\ge\; p_0 + (1-p_0)\P(M_N\ge x\mid Y=0),
\]
which rearranges to
\[
\P(M_N\ge x\mid Y=0)\;\le\;\Big(\frac{p_0}{1-p_0}\Big)\Big(\frac{1-x}{x}\Big).
\]
Taking complements gives the stated strict-threshold bound; the ordinary CDF bound follows because $\P(M_N\le x\mid Y=0)\ge\P(M_N<x\mid Y=0)$.
\end{proof}

\begin{proposition}[Discrete-time correction identity]
For $x\in(p_0,1)$,
\begin{align*}
\P(M_N\ge x\mid Y=0)
={}&\frac{p_0}{1-p_0}\cdot\frac{1-x}{x} \\
&-\frac{1}{x(1-p_0)}
  \E\!\big[(p_{\tau_x}-x)\mathbf 1\{\tau_x<N\}\big] \\
&-\frac{1-x}{x(1-p_0)}\,\P(\tau_x=N).
\end{align*}
\end{proposition}

\begin{proof}
Fix $x\in(p_0,1)$. Since $x<1$, the event $\{Y=1\}$ is contained in $\{\tau_x\le N\}$, so
\[
\P(\tau_x\le N)=\P(Y=1)+(1-p_0)\P(M_N\ge x\mid Y=0).
\]
Optional stopping at $\tau_x\wedge N$ gives
\[
p_0=\E[p_{\tau_x\wedge N}]
=\E[p_{\tau_x}\mathbf 1\{\tau_x\le N\}]
=x\,\P(\tau_x\le N)+\E\!\big[(p_{\tau_x}-x)\mathbf 1\{\tau_x\le N\}\big].
\]
Combining the two displays and rearranging yields
\[
\P(M_N\ge x\mid Y=0)
=\frac{p_0}{1-p_0}\cdot\frac{1-x}{x}
-\frac{1}{x(1-p_0)}\,\E\!\big[(p_{\tau_x}-x)\mathbf 1\{\tau_x\le N\}\big].
\]
Finally,
\[
\E\!\big[(p_{\tau_x}-x)\mathbf 1\{\tau_x\le N\}\big]
=\E\!\big[(p_{\tau_x}-x)\mathbf 1\{\tau_x< N\}\big]+(1-x)\P(\tau_x=N),
\]
because $\tau_x=N$ implies $p_{\tau_x}=p_N=1$.
Substituting gives the stated decomposition.
\end{proof}

\section{Proofs of Continuous-Path Results} \label{app:continuous-proofs}

Let $(p_t)_{t\in[0,1]}$ be a bounded martingale with $p_0\in(0,1)$ and $p_1=Y\in\{0,1\}$, and assume path continuity.
Define $M=\sup_{t\in[0,1]}p_t$ and $\tau_x=\inf\{t\in[0,1]: p_t\ge x\}$.

\subsection{Unconditional distribution of \texorpdfstring{$M$}{M}}

\begin{theorem}[Continuous-path unconditional] \label{thm:supp-continuous-unconditional}
For $x\in[p_0,1)$,
\[
F_M(x)=\P(M\le x)=1-\frac{p_0}{x},
\]
and $F_M(x)=0$ for $x<p_0$, while $F_M(1)=1$.
In particular, $\P(M=1)=p_0$.
\end{theorem}

\begin{proof}
If $x<p_0$ then $M\ge p_0>x$ a.s.

Fix $x\in[p_0,1)$. Optional stopping at $\tau_x\wedge 1$ yields $\E[p_{\tau_x\wedge 1}]=p_0$.
By continuity, on $\{\tau_x<1\}$ we have $p_{\tau_x}=x$, and on $\{\tau_x\ge 1\}$ we have $p_1=Y$.
Thus
\[
p_0 = x\,\P(\tau_x<1) + \E[Y\mathbf 1\{\tau_x\ge 1\}].
\]
But if $Y=1$, continuity and $p_1=1>x$ imply that $p_t\ge x$ for some $t<1$, and hence $\tau_x<1$; therefore
$\E[Y\mathbf 1\{\tau_x\ge 1\}]=0$ for $x<1$.
Therefore $p_0 = x\,\P(\tau_x<1)$, i.e.\ $\P(M\ge x)=\P(\tau_x<1)=p_0/x$.
Because this tail is continuous in $x$ on $[p_0,1)$, $M$ has no atoms there.
Thus $\P(M\le x)=\P(M<x)=1-p_0/x$ on that interval.
Finally, $M=1$ iff $Y=1$, so $\P(M=1)=p_0$.
\end{proof}

\subsection{Conditional distribution of \texorpdfstring{$M$ given $Y=0$}{M given Y=0}}

\begin{theorem}[Continuous-path conditional] \label{thm:supp-continuous-conditional}
For $x\in[p_0,1)$,
\[
F_{M\mid Y=0}(x)=\P(M\le x\mid Y=0)
=
1-\Big(\frac{p_0}{1-p_0}\Big)\Big(\frac{1-x}{x}\Big),
\]
and $F_{M\mid Y=0}(x)=0$ for $x<p_0$, while $F_{M\mid Y=0}(1)=1$.
\end{theorem}

\begin{proof}
Fix $x\in[p_0,1)$. As before,
\[
\P(M\ge x)=\P(Y=1)+(1-p_0)\P(M\ge x\mid Y=0)=p_0+(1-p_0)\P(M\ge x\mid Y=0).
\]
By Theorem~\ref{thm:supp-continuous-unconditional}, $\P(M\ge x)=p_0/x$, hence
\[
\frac{p_0}{x}=p_0+(1-p_0)\P(M\ge x\mid Y=0)
\quad\Rightarrow\quad
\P(M\ge x\mid Y=0)=\Big(\frac{p_0}{1-p_0}\Big)\Big(\frac{1-x}{x}\Big).
\]
This conditional tail is continuous in $x$ below one, so it has no atoms there; taking complements yields the stated CDF.
Finally, on $\{Y=0\}$ the process cannot attain $1$ (if $p_t=1$ then $\P(Y=1\mid\mathcal F_t)=1$), so $M<1$ a.s.\ and $F_{M\mid Y=0}(1)=1$.
\end{proof}

\section{Distribution of \texorpdfstring{$M_\lambda$}{M\_lambda} (Two Teams)} \label{app:two-player-derivation}

Let $(p_t)$ and $(q_t)$ be the two teams' win-probability martingales, where $q_t = 1 - p_t$ and $p_0 = \P(Y=1)$; assume $p_0 \ge 0.5$ without loss of generality.
The maximum win probability attained by the eventual loser is
\begin{equation} \label{eq:supp-M-lambda-def}
    M_\lambda \equiv \sup_{0 \leq t \leq 1} \Big( p_t \boldsymbol{1}\{Y = 0\} + q_t \boldsymbol{1}\{Y = 1\} \Big).
\end{equation}
By the law of total probability,
\begin{align} \label{eq:supp-M-lambda-total-prob}
    F_{M_\lambda}(x) = \P(M_\lambda \le x) &= \P(M_\lambda \le x \mid Y=0) (1-p_0) + \P(M_\lambda \le x \mid Y=1) p_0.
\end{align}

When team~A loses, Theorem~\ref{thm:supp-continuous-conditional} gives
\begin{align}
    F_{M_\lambda \mid Y=0}(x) = \P(M_\lambda \le x \mid Y = 0) &=
    \begin{cases}
        0 & \text{for } x < p_0, \\
        1 - \left(\frac{p_0}{1-p_0}\right) \left(\frac{1-x}{x}\right) & \text{for } x \in [p_0,1), \\
        1 & \text{for } x = 1.
    \end{cases} \label{eq:supp-M-lambda-A}
\end{align}
Applying the same formula with $p_0$ replaced by $1-p_0$ gives
\begin{align}
    F_{M_\lambda \mid Y=1}(x) = \P(M_\lambda \le x \mid Y = 1) &=
    \begin{cases}
        0 & \text{for } x < 1-p_0, \\
        1 - \left(\frac{1-p_0}{p_0}\right) \left(\frac{1-x}{x}\right) & \text{for } x \in [1-p_0,1), \\
        1 & \text{for } x = 1.
    \end{cases} \label{eq:supp-M-lambda-B}
\end{align}

With $p_0 \ge 0.5$ (team~A favored), substituting \eqref{eq:supp-M-lambda-A} and \eqref{eq:supp-M-lambda-B} into \eqref{eq:supp-M-lambda-total-prob} gives three regions:
\begin{enumerate}[leftmargin=*]
    \item For $0 \le x < 1-p_0$, neither \eqref{eq:supp-M-lambda-A} nor \eqref{eq:supp-M-lambda-B} contributes, so $F_{M_\lambda}(x) = 0$.

    \item For $1-p_0 \le x < p_0$, only \eqref{eq:supp-M-lambda-B} contributes:
    \begin{align*}
        F_{M_\lambda}(x) &= (1-p_0) \cdot 0 + p_0 \cdot \left[1 - \left(\frac{1-p_0}{p_0}\right) \left(\frac{1-x}{x}\right)\right] \\
        &= p_0 - (1-p_0)\left(\frac{1-x}{x}\right) = 1 - \frac{1-p_0}{x}.
    \end{align*}

    \item For $x \ge p_0$, both \eqref{eq:supp-M-lambda-A} and \eqref{eq:supp-M-lambda-B} contribute:
    \begin{align*}
        F_{M_\lambda}(x) &= (1-p_0)\left[1 - \left(\frac{p_0}{1-p_0}\right) \left(\frac{1-x}{x}\right)\right] + p_0\left[1 - \left(\frac{1-p_0}{p_0}\right) \left(\frac{1-x}{x}\right)\right] \\
        &= \left[(1-p_0) - p_0\left(\frac{1-x}{x}\right)\right] + \left[p_0 - (1-p_0)\left(\frac{1-x}{x}\right)\right] \\
        &= 1 - \frac{1-x}{x} = \frac{2x-1}{x} = 2 - \frac{1}{x}.
    \end{align*}
\end{enumerate}

Combining the three regions gives the piecewise cumulative distribution function
\begin{equation*}
    F_{M_\lambda}(x) = \P(M_\lambda \le x) = \begin{cases}
        0 & \text{for } 0 \le x < 1-p_0, \\[0.5em]
        1 - \frac{1-p_0}{x} & \text{for } 1-p_0 \le x < p_0, \\[0.5em]
        2 - \frac{1}{x} & \text{for } p_0 \le x < 1, \\[0.5em]
        1 & \text{for } x = 1.
    \end{cases}
\end{equation*}

\section{Numerical Benchmarks} \label{sec:supp-benchmarks}

The piecewise CDF in Supplementary Section~\ref{app:two-player-derivation} gives the selected values in Table~\ref{tab:supp-numerical-benchmarks}.
When the peak threshold satisfies $x\ge p_0$, the tail is universal:
\[
\P(M_\lambda\ge x)=\frac{1}{x}-1.
\]
In the intermediate region, $1-p_0\le x<p_0$, the corresponding probability is $(1-p_0)/x$.
The first three rows follow a game with $p_0=2/3$ across the boundary between these regions; the remaining rows give the practically important upper-tail thresholds.
For discretely reported paths, these continuous-path probabilities are conservative upper bounds.

\begin{table}[!tbp]
\centering
\small
\caption{Selected benchmark probabilities for the eventual loser's peak.}
\label{tab:supp-numerical-benchmarks}
\begin{tabular}{lccc}
\hline
Case & Starting favorite & Peak threshold $x$ & $\P(M_\lambda\ge x)$ \\
\hline
Intermediate region & $p_0=2/3$ & $1/2$ & $2/3$ \\
Universal boundary & $p_0=2/3$ & $2/3$ & $1/2$ \\
Universal tail & $p_0=2/3$ & $3/4$ & $1/3$ \\
Universal tail & $p_0\le 0.90$ & $0.90$ & $1/9\;(0.111)$ \\
Universal tail & $p_0\le 0.95$ & $0.95$ & $1/19\;(0.0526)$ \\
Universal tail & $p_0\le 0.99$ & $0.99$ & $1/99\;(0.0101)$ \\
\hline
\end{tabular}
\end{table}

\section{Dependence Validity and Bootstrap} \label{sec:supp-inference}

Intervals and $p$-values in Sections~\ref{sec:supp-inference}--\ref{sec:supp-results} use $B=\DyadicBootstrapReplicates$ season-stratified dyadic draws with seed \DyadicBootstrapSeed.

\subsection{Exchangeable-array specialization}

The empirical-process bootstrap theorem of \citet{DaveziesDHAultfoeuilleGuyonvarch2021} and its multiple-observations-per-dyad extension cover zero or repeated sports matches between a pair.
The following corollary specializes those results to an equal-game empirical CDF with within-season normalization.

For season $s$, let $n_s$ be the number of teams, $N_{s,ab}$ the number of games between teams $a<b$, and $U_{s,ab\ell}$ the benchmark score in meeting $\ell$.
For $h_t(u)=\mathbf 1\{u<t\}$, define the empirical numerator and denominator
\[
\widehat A_s(t)
=\binom{n_s}{2}^{-1}
  \sum_{a<b}\sum_{\ell=1}^{N_{s,ab}}h_t(U_{s,ab\ell}),
\qquad
\widehat B_s
=\binom{n_s}{2}^{-1}\sum_{a<b}N_{s,ab}.
\]
The strict empirical CDF among games in season $s$ is therefore
\[
\widehat F_s(t-)=\frac{\widehat A_s(t)}{\widehat B_s}.
\]
Its population counterpart is $F_s(t-)=A_s(t)/B_s$, where
\[
A_s(t)=\E\!\left[\sum_{\ell=1}^{N_{s,12}}h_t(U_{s,12\ell})\right],
\qquad
B_s=\E[N_{s,12}].
\]
Let $\omega_{s,n}\ge0$ be fixed stratification weights, with $\sum_s\omega_{s,n}=1$ and $\omega_{s,n}\to\omega_s$.
The application conditions on the observed season composition and sets these weights to each season's share of analyzed games.
Thus
\[
\widehat F(t-)=\sum_{s=1}^S\omega_{s,n}\widehat F_s(t-),
\qquad
F_{\omega,n}(t-)=\sum_{s=1}^S\omega_{s,n}F_s(t-).
\]

Within season $s$, let $(W_{s1},\ldots,W_{sn_s})$ be multinomial team multiplicities from $n_s$ draws.
The bootstrap analogues are
\begin{align*}
\widehat A_s^*(t)
&=\binom{n_s}{2}^{-1}
  \sum_{a<b}W_{sa}W_{sb}
  \sum_{\ell=1}^{N_{s,ab}}h_t(U_{s,ab\ell}),\\
\widehat B_s^*
&=\binom{n_s}{2}^{-1}
  \sum_{a<b}W_{sa}W_{sb}N_{s,ab},
\qquad
\widehat F_s^*(t-)=\frac{\widehat A_s^*(t)}{\widehat B_s^*},
\end{align*}
and $\widehat F^*(t-)=\sum_s\omega_{s,n}\widehat F_s^*(t-)$.
This ratio is exactly the CDF obtained by normalizing the raw game weights $W_{sa}W_{sb}$ within season.

\begin{corollary}[CDF specialization of the exchangeable-array bootstrap] \label{cor:dyadic-cdf}
Let the number of seasons $S$ be fixed, $n_{\min}=\min_s n_s\to\infty$, and $n_{\min}/n_s\to\kappa_s\in(0,1]$.
Within each season, suppose
\[
\bigl(N_{s,ab},(U_{s,ab\ell})_{\ell\ge1}\bigr)_{a<b}
\]
is jointly exchangeable and dissociated under team relabeling, and suppose the season arrays are independent.
Assume $N_{s,ab}\le C<\infty$ almost surely and $B_s>0$.
Assume also that the first-order dyadic projection of the normalized CDF process is nondegenerate, meaning that its limiting covariance function is not identically zero.
Then, in $\ell^\infty([0,1])$,
\[
\sqrt{n_{\min}}\{\widehat F(t-)-F_{\omega,n}(t-)\}
\rightsquigarrow \mathbb G(t),
\]
and, conditionally on the observed arrays,
\[
\sqrt{n_{\min}}\{\widehat F^*(t-)-\widehat F(t-)\}
\rightsquigarrow \mathbb G(t).
\]
Consequently, at continuity points of the limiting supremum law, the conditional distribution of
\[
\sqrt{n_{\min}}\sup_t\{\widehat F(t-)-\widehat F^*(t-)\}
\]
consistently estimates the distribution of
\[
\sqrt{n_{\min}}\sup_t\{F_{\omega,n}(t-)-\widehat F(t-)\}.
\]
\end{corollary}

\begin{proof}
The class
\[
\mathcal H_+=\{u\mapsto\mathbf 1(u<t):t\in[0,1]\}\cup\{1\}
\]
is a pointwise measurable VC class with bounded envelope: rational thresholds form a countable pointwise-dense subclass.
Parts 2 and 3 of Supplemental Theorem S1 in \citet{DaveziesDHAultfoeuilleGuyonvarch2021} therefore give joint weak convergence of
$\sqrt{n_s}(\widehat A_s-A_s,\widehat B_s-B_s)$ and its multinomial-team bootstrap counterpart.
The bounded cell count supplies the extension's moment conditions.
An unordered symmetric dyad can be embedded in the theorem's ordered-array notation by assigning the same cell to $(a,b)$ and $(b,a)$; the common factor of two cancels from the CDF ratio.

Because $B_s>0$, the map $\Psi(A,B)=A/B$ is Hadamard differentiable, with derivative
\[
\dot\Psi_{(A_s,B_s)}(\alpha,\beta)
=\frac{\alpha}{B_s}-\frac{A_s\beta}{B_s^2}.
\]
The bootstrap functional delta method therefore gives the two process limits for each normalized season CDF.
The second term carries the denominator fluctuation into the normalized process.
A finite independent weighted sum, together with $n_{\min}/n_s\to\kappa_s$, gives the pooled limits.
Finally, $f\mapsto\sup_t\{-f(t)\}$ is Lipschitz on $\ell^\infty([0,1])$, which gives the last assertion directly, without a symmetry argument.
\end{proof}

\begin{corollary}[Conservative upper-tail test] \label{cor:conservative-test}
Suppose in addition that the discrete-feed null satisfies $F_{\omega,n}(t-)\ge t$ for every $t\in[0,1]$.
Let $c_{1-\alpha}^*$ be the conditional $(1-\alpha)$ quantile of the scaled recentered bootstrap statistic in Corollary~\ref{cor:dyadic-cdf}.
Define
\[
D_n^{\mathrm{upper}}=\sup_t\{t-\widehat F(t-)\}
\]
and reject when $\sqrt{n_{\min}}D_n^{\mathrm{upper}}>c_{1-\alpha}^*$.
This test has asymptotic size at most $\alpha$ whenever the limiting supremum distribution is continuous at its $(1-\alpha)$ quantile.
\end{corollary}

\begin{proof}
Under the null,
\[
t-\widehat F(t-)
\le F_{\omega,n}(t-)-\widehat F(t-)
\]
pointwise.
The result follows by taking suprema and applying Corollary~\ref{cor:dyadic-cdf}.
\end{proof}

Corollaries~\ref{cor:dyadic-cdf}--\ref{cor:conservative-test} use a dense-array asymptotic and a nondegenerate first-order dyadic projection.
A vanishing projection can produce a second-order limit and conservative pigeonhole resampling; more generally, \citet{Menzel2021} rules out uniform bootstrap consistency across all multiway-dependence regimes.
Professional schedules have fixed conference and division structures, and franchises persist across seasons.
Section~\ref{sec:supp-validation} evaluates finite-sample size on the observed schedules, and the franchise-linked specification in Table~\ref{tab:supp-specification} carries team effects across seasons.

\subsection{Season-stratified dyadic pigeonhole bootstrap}

Write game $g$ in season $s$ as the dyad $(a_g,b_g)$.
The resampling algorithm is:
\begin{enumerate}[leftmargin=2em]
\item Within each season, sample its observed teams multinomially with replacement and record team multiplicities $W_{sa}$.
\item Give game $g$ raw weight $W_{s a_g}W_{s b_g}$. Repeat meetings between the same teams therefore receive the same endpoint weight. If a season draw assigns zero total weight to its observed schedule, redraw that season.
\item Normalize the positive game weights within season to sum to that season's observed number of contributions. This retains the empirical season composition while allowing arbitrary dependence among games that share a team.
\item Reuse the same team multiplicities for the PIT process, tail rates, first-crossing residuals, and fixed-clock calibration cells.
\end{enumerate}
This applies Owen's pigeonhole principle---resampling indexing units rather than treating cells as independent---to a symmetric dyadic schedule \citep{Owen2007}.
The exchangeable-array empirical-process results of \citet{DaveziesDHAultfoeuilleGuyonvarch2021} cover recentered empirical CDF and Kolmogorov--Smirnov processes for dyads, including repeated observations between the same pair.

For tied PIT scores $u_1,\ldots,u_n$, the observed upper statistic is computed exactly as
\[
D_n^{\mathrm{upper}}
=\max_{u\in\{u_1,\ldots,u_n\}}\{u-\hat F_U(u-)\}.
\]
For bootstrap draw $b$, let $\hat F_{U,b}^*$ be the season-normalized weighted CDF and calculate
\[
D_{n,b}^*=\sup_t\{\hat F_U(t)-\hat F_{U,b}^*(t)\}.
\]
For the common observed support, this right-continuous supremum equals the strict-CDF supremum in Corollary~\ref{cor:dyadic-cdf}: both empirical CDFs have the same total mass, so their right-limit and left-limit differences run through the same values, with zero at opposite endpoints.
The finite-bootstrap $p$-value is $(1+\sum_b\mathbf 1\{D_{n,b}^*\ge D_n^{\mathrm{upper}}\})/(B+1)$.

\section{Published Feeds and Fixed-Time Calibration} \label{sec:supp-data}

Both applications contain completed regular-season games only.
NFL seasons are labeled by starting year and include January games from the following calendar year; NBA seasons are labeled by ending year.
Because ESPN's NBA season-type field also includes All-Star exhibitions, both participants must be NBA franchises; this excludes \NBAExcludedExhibitionEvents{} All-Star or Rising Stars events, including \NBAExcludedExhibitionFeeds{} with usable probability paths.
Table~\ref{tab:supp-coverage} gives the complete acquisition ledger.
``Eligible'' means completed and non-tied, while ``analyzed'' additionally requires a usable public probability path.

\begin{table}[!tbp]
\centering
\small
\caption{Regular-season schedule and public-feed coverage by season.}
\label{tab:supp-coverage}
\textbf{NFL}\\[0.3em]
\begin{tabular}{lrrrr}
\hline
Season & Scheduled & Eligible & Analyzed & Missing \\
\hline
\NFLCoverageRows
\hline
\end{tabular}
\par\vspace{0.8em}
\textbf{NBA}\\[0.3em]
\begin{tabular}{lrrrr}
\hline
Season & Scheduled & Eligible & Analyzed & Missing \\
\hline
\NBACoverageRows
\hline
\end{tabular}
\end{table}

Each application retains its published resolution.
The NFL paths contain \NFLPublishedUpdates{} updates, with median \NFLMedianUpdatesPerGame{} per game (interquartile range \NFLUpdateQOne--\NFLUpdateQThree); the NBA paths contain \NBAPublishedUpdates{}, with median \NBAMedianUpdatesPerGame{} (\NBAUpdateQOne--\NBAUpdateQThree).

\FloatBarrier
\subsection{Yeh-style fixed-clock comparison}

Following the calibration-surface motivation of \citet{YehRiceDubin2022}, every retained home-team forecast is evaluated at $t=0,.05,\ldots,.95$ of regulation.
Linear interpolation is primary and last observation carried forward (LOCF) is the sensitivity.
At each time and for each method, empirical probability quantiles define at most ten bins; a repeated forecast value is never split between bins.
Every complete game therefore contributes exactly once to every time point.
Overtime updates are omitted from this common-clock comparison but remain in the collapse statistic.

\begin{figure}[!tbp]
\centering
\begin{minipage}[t]{0.49\linewidth}
  \centering
  \includegraphics[width=\linewidth]{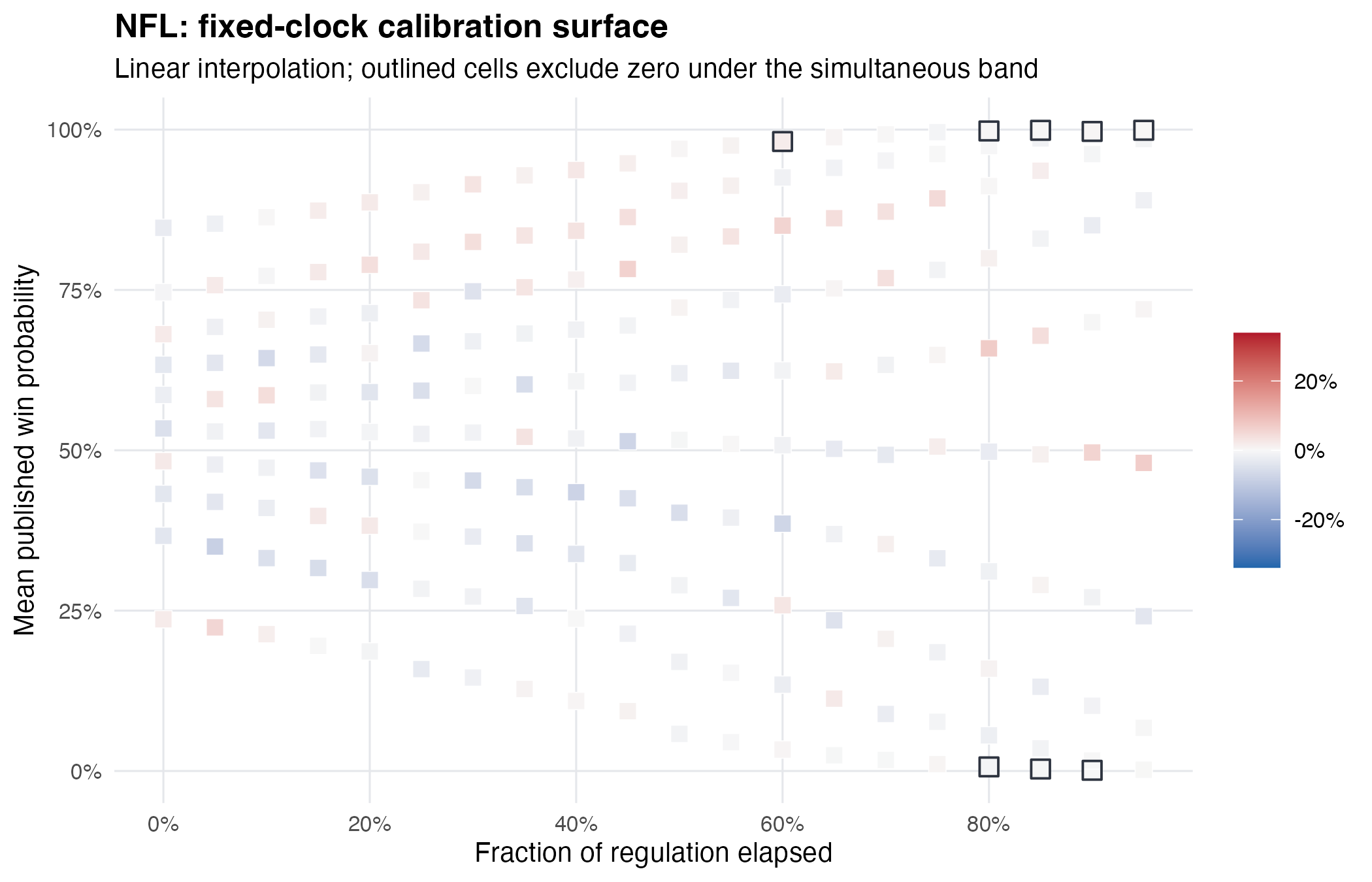}
  \\[-0.2em]{\small (a) NFL, linear interpolation}
\end{minipage}\hfill
\begin{minipage}[t]{0.49\linewidth}
  \centering
  \includegraphics[width=\linewidth]{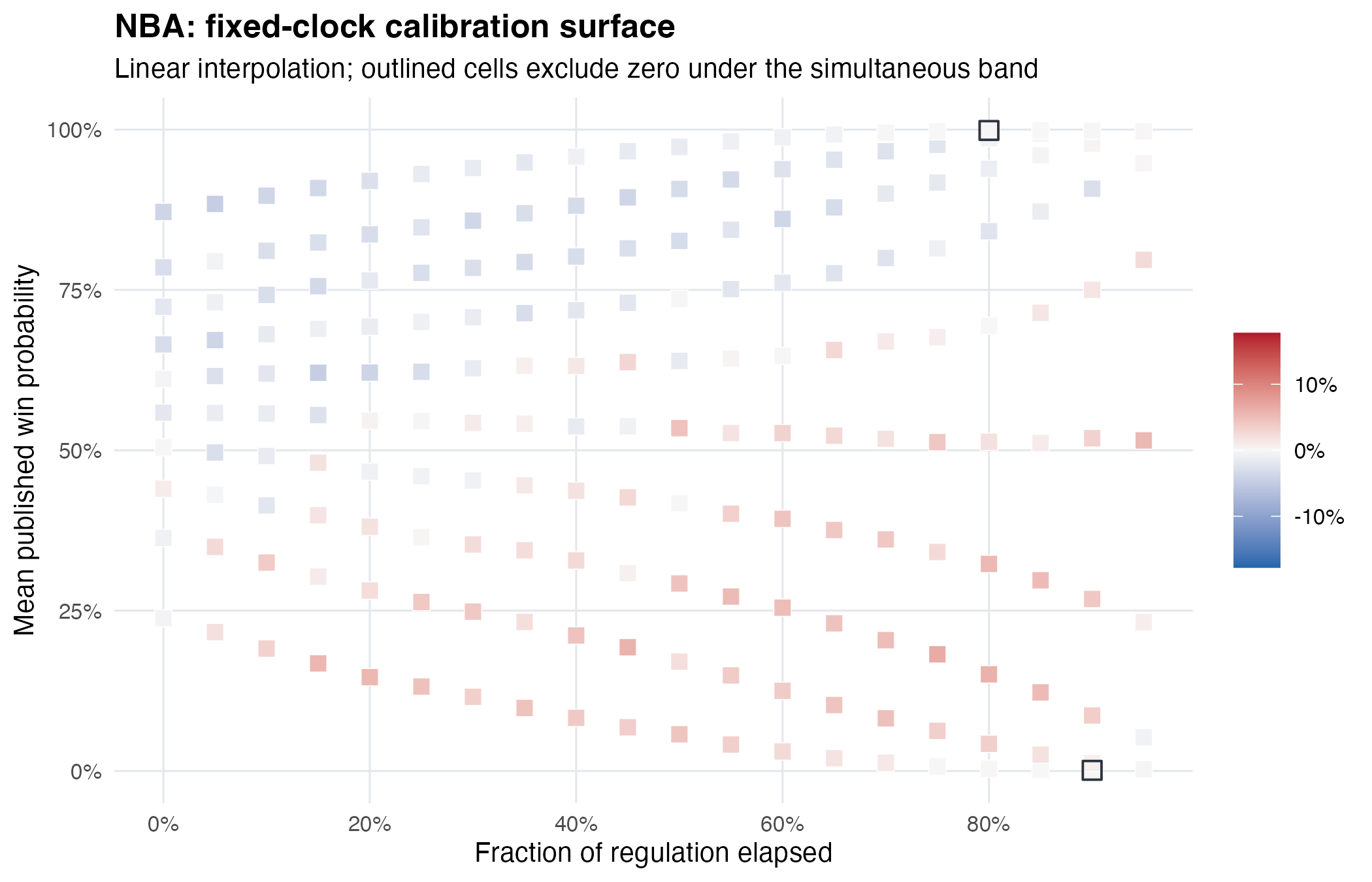}
  \\[-0.2em]{\small (b) NBA, linear interpolation}
\end{minipage}\\[0.8em]
\begin{minipage}[t]{0.49\linewidth}
  \centering
  \includegraphics[width=\linewidth]{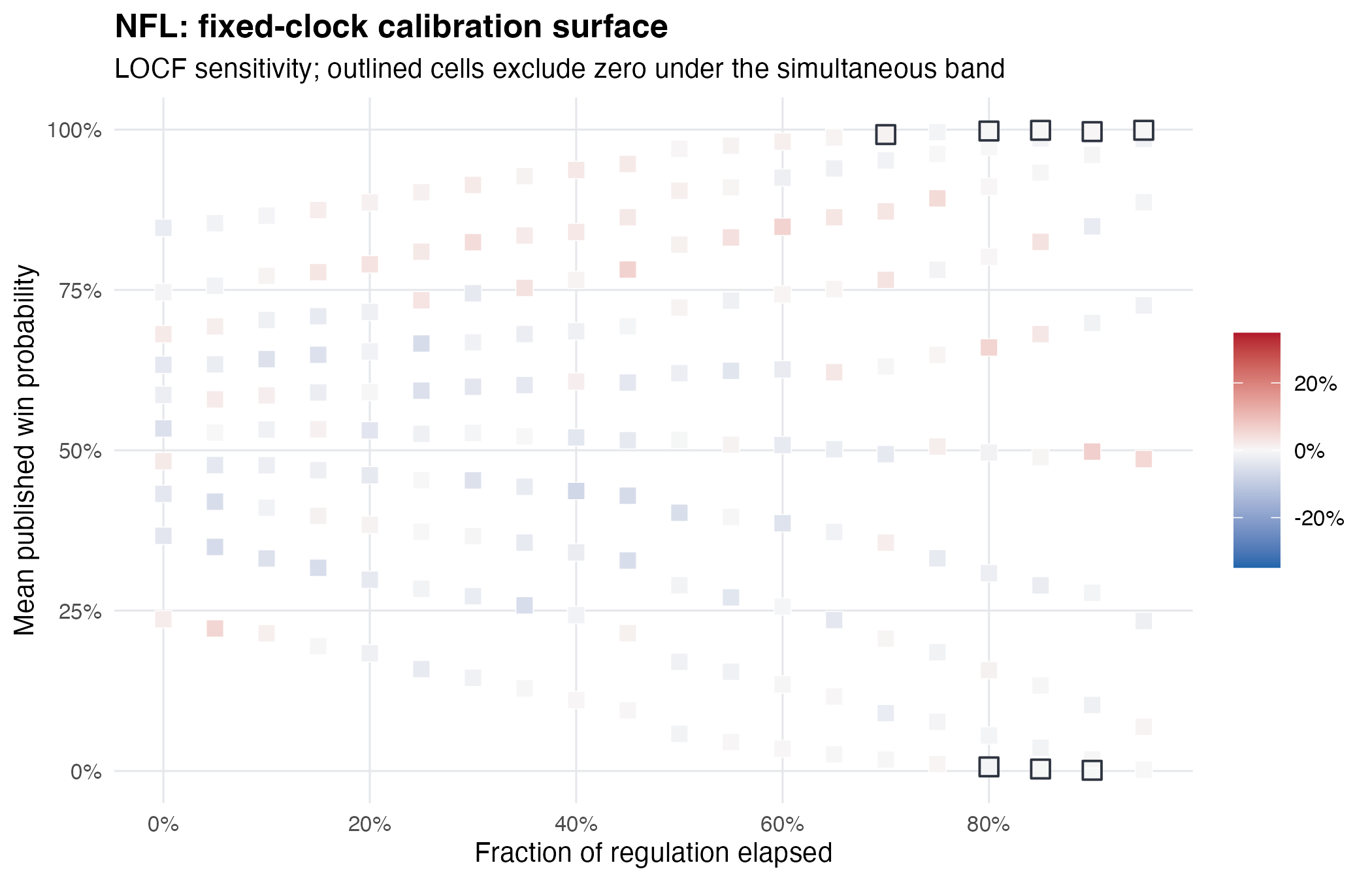}
  \\[-0.2em]{\small (c) NFL, LOCF}
\end{minipage}\hfill
\begin{minipage}[t]{0.49\linewidth}
  \centering
  \includegraphics[width=\linewidth]{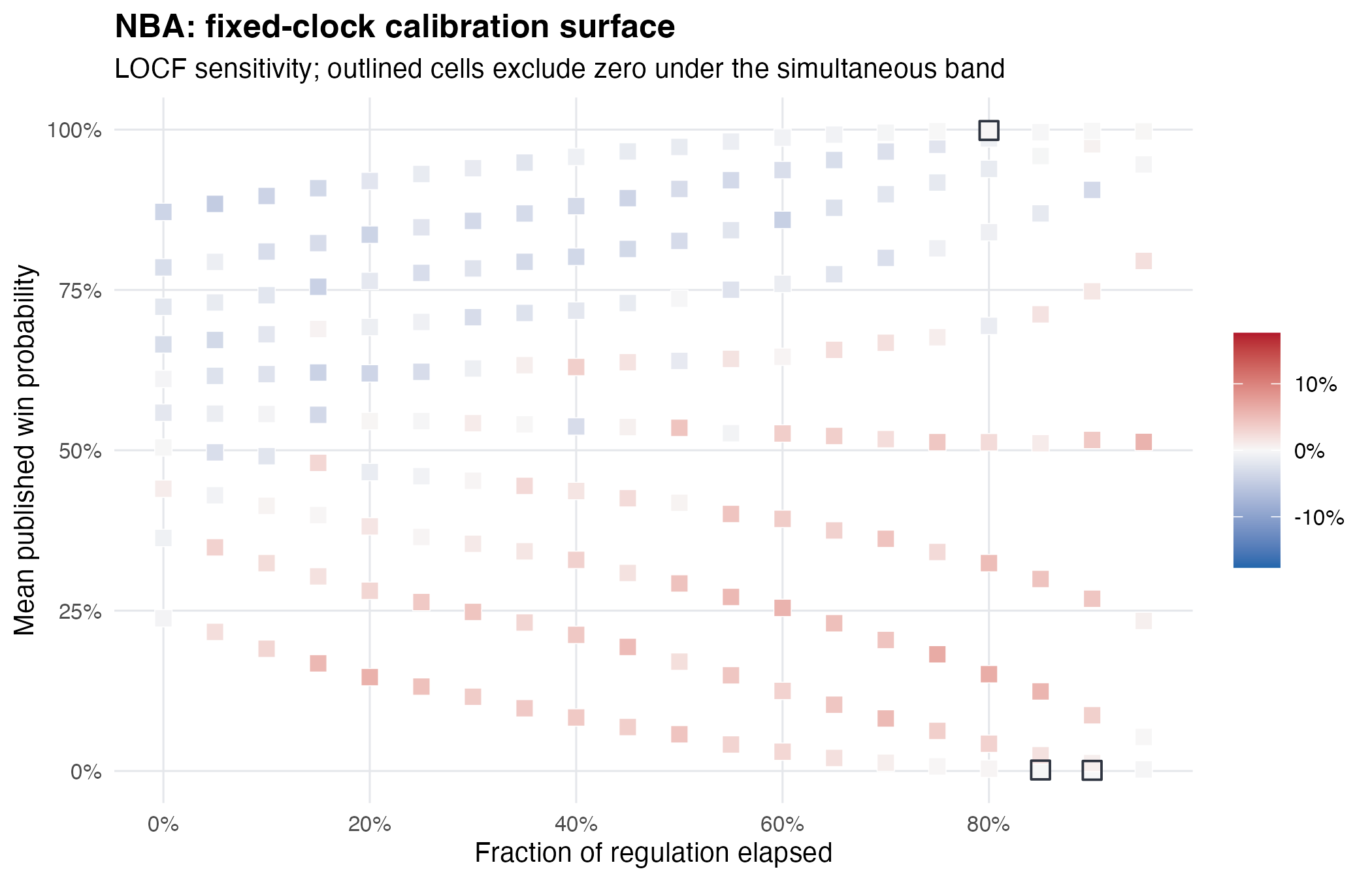}
  \\[-0.2em]{\small (d) NBA, LOCF}
\end{minipage}
\caption{Fixed-clock calibration surfaces. Color is observed win rate minus mean published probability; outlined cells exclude zero under the simultaneous dyadic band. The bin locations adapt to the forecast distribution at each time.}
\label{fig:supp-fixed-clock}
\end{figure}

The linear and LOCF analyses use the same \NFLFixedClockGames{} NFL games and \NBAFixedClockGames{} NBA games.
All NFL games have a complete common clock; \NBAFixedClockExcluded{} NBA paths are incomplete by $t=.95$ and are excluded from this comparison only.
For each time--probability cell, the observed-minus-forecast gap is resampled using the common dyadic draws.
A simultaneous band is obtained from the 95th percentile of the maximum absolute studentized gap over all cells.
Under linear interpolation, the simultaneous bands exclude zero in \NFLFixedClockLinearSignificantCells{} of \NFLFixedClockLinearCells{} NFL cells and \NBAFixedClockLinearSignificantCells{} of \NBAFixedClockLinearCells{} NBA cells; LOCF gives \NFLFixedClockLOCFSignificantCells{} of \NFLFixedClockLOCFCells{} and \NBAFixedClockLOCFSignificantCells{} of \NBAFixedClockLOCFCells{}, respectively.
The flagged cells lie almost entirely at or after 80\% of regulation and in the extreme probability bins.
In each flagged cell, the observed win rate is exactly zero or one while the corresponding mean forecast remains just inside the boundary.
Thus the fixed-clock analysis finds sparse late-game boundary discrepancies, while the interior cells remain within the simultaneous bands.
The descriptive root-mean-square cell gaps are \NFLFixedClockLinearRMS{} (linear) and \NFLFixedClockLOCFRMS{} (LOCF) in the NFL, and \NBAFixedClockLinearRMS{} and \NBAFixedClockLOCFRMS{} in the NBA.
Figure~\ref{fig:supp-fixed-clock} displays the full surfaces.

The fixed-clock surfaces evaluate reliability at prespecified times; the collapse benchmark evaluates the selected losing path.

\section{Full Collapse Results and Robustness} \label{sec:supp-results}

\subsection{Full tail and first-crossing results}

Table~\ref{tab:supp-tail} reports upper-tail rates at $0.90$, $0.95$, and $0.99$.
The NBA excess is positive at $0.90$ and $0.95$; its $0.99$ interval includes zero.

\begin{table}[!tbp]
\centering
\small
\caption{Upper-tail rates and excess over the continuous benchmark. Intervals are two-sided percentile dyadic-bootstrap intervals for the excess.}
\label{tab:supp-tail}
\begin{tabular}{lcccc}
\hline
League & Threshold & Tail rate & Benchmark rate & Excess (95\% CI) \\
\hline
NFL & 0.90 & \NFLTailNinetyRate & 10.0\% & \NFLTailNinetyExcess{} [\NFLTailNinetyCILower, \NFLTailNinetyCIUpper] \\
NFL & 0.95 & \NFLTailNinetyFiveRate & 5.0\% & \NFLTailNinetyFiveExcess{} [\NFLTailNinetyFiveCILower, \NFLTailNinetyFiveCIUpper] \\
NFL & 0.99 & \NFLTailNinetyNineRate & 1.0\% & \NFLTailNinetyNineExcess{} [\NFLTailNinetyNineCILower, \NFLTailNinetyNineCIUpper] \\
NBA & 0.90 & \NBATailNinetyRate & 10.0\% & \NBATailNinetyExcess{} [\NBATailNinetyCILower, \NBATailNinetyCIUpper] \\
NBA & 0.95 & \NBATailNinetyFiveRate & 5.0\% & \NBATailNinetyFiveExcess{} [\NBATailNinetyFiveCILower, \NBATailNinetyFiveCIUpper] \\
NBA & 0.99 & \NBATailNinetyNineRate & 1.0\% & \NBATailNinetyNineExcess{} [\NBATailNinetyNineCILower, \NBATailNinetyNineCIUpper] \\
\hline
\end{tabular}
\end{table}

For a team-side path, let $\tau_q$ be the first published update at or above $q$ with positive game time remaining.
If no preterminal crossing occurs, set $\tau_q=N$ and retain the indicator of crossing.
Because $\tau_q$ is bounded and $\{\tau_q<N\}\in\mathcal F_{\tau_q}$, sequential calibration implies
\[
\E\{(p_{\tau_q}-Y)\mathbf 1(\tau_q<N)\}=0.
\]
Optional stopping therefore centers the first-crossing residual at zero under sequential calibration.
Table~\ref{tab:supp-crossing} reports $\overline{1-Y}$, $\overline{1-p_{\tau_q}}$, and their difference $\bar R=\overline{p_{\tau_q}-Y}$.
The estimand is an average over crossing episodes; when two episodes come from the same game, they receive the same dyadic game weight.
The crossing estimand uses a published probability $p\ge q$; the tail estimand in Table~\ref{tab:supp-tail} uses the game-level benchmark score $U\ge q$.

\begin{table}[!tbp]
\centering
\small
\setlength{\tabcolsep}{4pt}
\caption{Preterminal first crossings. The interval is for the mean residual $\bar R$.}
\label{tab:supp-crossing}
\begin{tabular}{lcrrrrr}
\hline
League & $q$ & Episodes & Games & Observed loss & Implied loss & $\bar R$ (95\% CI) \\
\hline
NFL & 0.90 & \NFLCrossingNinetyEpisodes & \NFLCrossingNinetyGames & \NFLCrossingNinetyLoss & \NFLCrossingNinetyImplied & \NFLCrossingNinetyResidual{} [\NFLCrossingNinetyCILower, \NFLCrossingNinetyCIUpper] \\
NFL & 0.95 & \NFLCrossingNinetyFiveEpisodes & \NFLCrossingNinetyFiveGames & \NFLCrossingNinetyFiveLoss & \NFLCrossingNinetyFiveImplied & \NFLCrossingNinetyFiveResidual{} [\NFLCrossingNinetyFiveCILower, \NFLCrossingNinetyFiveCIUpper] \\
NFL & 0.99 & \NFLCrossingNinetyNineEpisodes & \NFLCrossingNinetyNineGames & \NFLCrossingNinetyNineLoss & \NFLCrossingNinetyNineImplied & \NFLCrossingNinetyNineResidual{} [\NFLCrossingNinetyNineCILower, \NFLCrossingNinetyNineCIUpper] \\
NBA & 0.90 & \NBACrossingNinetyEpisodes & \NBACrossingNinetyGames & \NBACrossingNinetyLoss & \NBACrossingNinetyImplied & \NBACrossingNinetyResidual{} [\NBACrossingNinetyCILower, \NBACrossingNinetyCIUpper] \\
NBA & 0.95 & \NBACrossingNinetyFiveEpisodes & \NBACrossingNinetyFiveGames & \NBACrossingNinetyFiveLoss & \NBACrossingNinetyFiveImplied & \NBACrossingNinetyFiveResidual{} [\NBACrossingNinetyFiveCILower, \NBACrossingNinetyFiveCIUpper] \\
NBA & 0.99 & \NBACrossingNinetyNineEpisodes & \NBACrossingNinetyNineGames & \NBACrossingNinetyNineLoss & \NBACrossingNinetyNineImplied & \NBACrossingNinetyNineResidual{} [\NBACrossingNinetyNineCILower, \NBACrossingNinetyNineCIUpper] \\
\hline
\end{tabular}
\end{table}

\subsection{Feed and specification sensitivity}

Table~\ref{tab:supp-specification} reports sensitivity to feed corrections, rounding, dependence structure, threshold convention, and time sampling.
The crossing rows report the $0.95$ residual, while the fixed-clock rows report descriptive root-mean-square calibration gaps over time--probability cells.
For reported probabilities rounded to three decimals, lower and upper PIT vectors are obtained by allowing each probability to range within $\pm0.0005$ and using monotonicity of the benchmark transform.
The dyad-plus-calendar row multiplies the team multiplicity by a circular two-week block multiplicity within each season, retaining team dependence while allowing common model-period shocks.
The franchise-linked row maps relocations to a common franchise and applies one team multiplicity across all seasons, allowing persistent team effects.

\begin{table}[!tbp]
\centering
\small
\setlength{\tabcolsep}{4pt}
\caption{Feed and specification sensitivity. Brackets denote 95\% intervals where shown.}
\label{tab:supp-specification}
\begin{tabular}{lll}
\hline
Specification & NFL & NBA \\
\hline
Global, corrected: $D$ ($p$) & \NFLGlobalD{} (\NFLGlobalDyadicP) & \NBAGlobalD{} (\NBAGlobalDyadicP) \\
Global, raw: $D$ ($p$) & \NFLRawGlobalD{} (\NFLRawGlobalP) & \NBARawGlobalD{} (\NBARawGlobalP) \\
Exclude patched games: $D$ ($p$) & \NFLExcludePatchedGlobalD{} (\NFLExcludePatchedGlobalP) & \NBAExcludePatchedGlobalD{} (\NBAExcludePatchedGlobalP) \\
Rounding bound for $D$ & [\NFLRoundingLowerGlobalD, \NFLRoundingUpperGlobalD] & [\NBARoundingLowerGlobalD, \NBARoundingUpperGlobalD] \\
Rounding bound for $\widehat\P(U\ge.95)$ & [\NFLRoundingLowerTailNinetyFiveRate, \NFLRoundingUpperTailNinetyFiveRate] & [\NBARoundingLowerTailNinetyFiveRate, \NBARoundingUpperTailNinetyFiveRate] \\
Dyad plus calendar: $D$ ($p$) & \NFLDyadicCalendarGlobalD{} (\NFLDyadicCalendarGlobalP) & \NBADyadicCalendarGlobalD{} (\NBADyadicCalendarGlobalP) \\
Franchise linked: $D$ ($p$) & \NFLFranchiseLinkedGlobalD{} (\NFLFranchiseLinkedGlobalP) & \NBAFranchiseLinkedGlobalD{} (\NBAFranchiseLinkedGlobalP) \\
Crossing, preterminal $p\ge.95$ & \NFLCrossingNinetyFiveResidual{} [\NFLCrossingNinetyFiveCILower, \NFLCrossingNinetyFiveCIUpper] & \NBACrossingNinetyFiveResidual{} [\NBACrossingNinetyFiveCILower, \NBACrossingNinetyFiveCIUpper] \\
Crossing, preterminal $p>.95$ & \NFLCrossingNinetyFiveStrictResidual{} [\NFLCrossingNinetyFiveStrictCILower, \NFLCrossingNinetyFiveStrictCIUpper] & \NBACrossingNinetyFiveStrictResidual{} [\NBACrossingNinetyFiveStrictCILower, \NBACrossingNinetyFiveStrictCIUpper] \\
Crossing, terminal-inclusive & \NFLCrossingNinetyFiveAllResidual{} [\NFLCrossingNinetyFiveAllCILower, \NFLCrossingNinetyFiveAllCIUpper] & \NBACrossingNinetyFiveAllResidual{} [\NBACrossingNinetyFiveAllCILower, \NBACrossingNinetyFiveAllCIUpper] \\
Fixed clock, linear RMS & \NFLFixedClockLinearRMS & \NBAFixedClockLinearRMS \\
Fixed clock, LOCF RMS & \NFLFixedClockLOCFRMS & \NBAFixedClockLOCFRMS \\
\hline
\end{tabular}
\end{table}

At $0.95$, strict versus weak tail counting differs only through the observed atom: the weak and strict NBA rates are \NBATailNinetyFiveRate{} and \NBATailNinetyFiveStrictRate{}, with \NBATailNinetyFiveAtomRate{} exactly at the boundary; the corresponding NFL values are \NFLTailNinetyFiveRate{}, \NFLTailNinetyFiveStrictRate{}, and \NFLTailNinetyFiveAtomRate{}.
The path audit identified \NBAPatchedGames{} NBA and \NFLPatchedGames{} NFL games containing loser-side exact-one feed values.
The raw, corrected, and exclusion rows in Table~\ref{tab:supp-specification} use identical team draws.

\subsection{Season persistence}

The acquisition ledger in Table~\ref{tab:supp-coverage} gives explicit missing-feed accounting for both leagues.
NFL coverage is complete; minimum NBA season coverage is \NBAMinimumFeedCoverage{}.

Table~\ref{tab:supp-loso} gives descriptive stability summaries after omitting one season at a time.
Across omissions, the NBA global statistic ranges from \NBALOSODMin{} to \NBALOSODMax{}, while the NFL statistic ranges from \NFLLOSODMin{} to \NFLLOSODMax{}.

\begin{table}[!tbp]
\centering
\small
\caption{Leave-one-season-out global statistic and $0.95$ tail excess.}
\label{tab:supp-loso}
\begin{minipage}[t]{0.47\linewidth}
\centering
\textbf{NFL}\\[0.3em]
\begin{tabular}{lrrr}
\hline
Omitted & Games & $D$ & Excess \\
\hline
\NFLLeaveSeasonRows
\hline
\end{tabular}
\end{minipage}\hfill
\begin{minipage}[t]{0.47\linewidth}
\centering
\textbf{NBA}\\[0.3em]
\begin{tabular}{lrrr}
\hline
Omitted & Games & $D$ & Excess \\
\hline
\NBALeaveSeasonRows
\hline
\end{tabular}
\end{minipage}
\end{table}

\FloatBarrier
Across specifications, the NBA departure remains a recurrent upper-tail feature.
The tail excess is positive at $0.90$ and $0.95$, first-crossing residuals are positive at all three thresholds, and the global test rejects for the raw paths, corrected paths, patched-game exclusion, calendar blocks, and franchise linkage.
Across the leave-one-season-out samples, $D_n^{\mathrm{upper}}$ ranges from \NBALOSODMin{} to \NBALOSODMax{} and the $0.95$ tail excess ranges from \NBALOSOTailMin{} to \NBALOSOTailMax{}.
The game-level $0.99$ tail interval includes zero, marking the limit of the upper-tail evidence.
Across specifications, the NFL retains the global conclusion of no detected departure; its positive $0.99$ first-crossing residual is a localized result.

\section{Finite-Schedule Null Simulation} \label{sec:supp-validation}

Finite-sample Type I error is evaluated while retaining the teams and matchups in each observed season schedule.
For game $g$ between teams $a_g$ and $b_g$, generate
\[
Z_g=\sqrt{\rho}\,(A_{s,a_g}+A_{s,b_g})/\sqrt{2}
    +\sqrt{1-\rho}\,E_g,
\qquad U_g=\Phi(Z_g),
\]
where team effects and game shocks are independent standard normals and team effects are redrawn by season.
Marginally, $Z_g$ is standard normal and $U_g$ is uniform, placing every simulation on the boundary of the continuous calibration null.
Games sharing a team are correlated through the team effect, with $\rho$ controlling the strength of that dependence.
For each $\rho\in\{0,0.25,0.5\}$, the global test is applied to \ValidationOuterSamples{} Monte Carlo samples using \ValidationInnerReplicates{} inner pigeonhole draws.
The proportion rejected estimates finite-schedule Type I error at nominal level 5\%.
When $\rho=0$, the pigeonhole procedure is expected to be conservative because the dyadic projection is degenerate; this is the regime in which multiway bootstrap behavior requires particular care \citep{Menzel2021}.
Table~\ref{tab:supp-validation} reports the resulting rejection rates.

\begin{table}[!tbp]
\centering
\caption{Empirical 5\% rejection rates in the finite-schedule null simulation (\ValidationOuterSamples{} Monte Carlo samples; \ValidationInnerReplicates{} inner draws).}
\label{tab:supp-validation}
\begin{tabular}{lcc}
\hline
League & $\rho$ & Dyadic rejection rate \\
\hline
NFL & 0 & \ValidationNFLRhoZeroDyadic \\
NFL & 0.25 & \ValidationNFLRhoTwentyFiveDyadic \\
NFL & 0.5 & \ValidationNFLRhoFiftyDyadic \\
NBA & 0 & \ValidationNBARhoZeroDyadic \\
NBA & 0.25 & \ValidationNBARhoTwentyFiveDyadic \\
NBA & 0.5 & \ValidationNBARhoFiftyDyadic \\
\hline
\end{tabular}
\end{table}
The test is conservative under these schedule-specific nulls.
Rejection rates rise with the shared-team component, from \ValidationNFLRhoZeroDyadic{} to \ValidationNFLRhoFiftyDyadic{} on the NFL schedule and from \ValidationNBARhoZeroDyadic{} to \ValidationNBARhoFiftyDyadic{} on the NBA schedule, remaining at or below the nominal 5\% level.
The simulation targets finite-schedule size under shared-team dependence; calendar and cross-season dependence are examined through the sensitivities in Section~\ref{sec:supp-results}.

\bibliographystyle{apalike}
\bibliography{references}

@article{Howard2020,
  author  = {Howard, Steven R. and Ramdas, Aaditya and McAuliffe, Jon and Sekhon, Jasjeet},
  title   = {{Time-Uniform Chernoff Bounds via Nonnegative Supermartingales}},
  journal = {Probability Surveys},
  volume  = {17},
  pages   = {257--317},
  year    = {2020},
  doi     = {10.1214/18-PS321}
}

@article{NikeghbaliYor2006,
  author  = {Nikeghbali, Ashkan and Yor, Marc},
  title   = {{Doob's Maximal Identity, Multiplicative Decompositions and Enlargements of Filtrations}},
  journal = {Illinois Journal of Mathematics},
  volume  = {50},
  number  = {1--4},
  pages   = {791--814},
  year    = {2006}
}

@manual{hoopR2023,
  author = {Gilani, Saiem},
  title  = {{hoopR: Access Men's Basketball Play by Play Data}},
  year   = {2023},
  note   = {R package version 2.1.0},
  url    = {https://CRAN.R-project.org/package=hoopR}
}

@manual{espnscrapeR2025,
  author = {Mock, Thomas},
  title  = {{espnscrapeR: Scrapes Or Collects NFL Data From ESPN}},
  year   = {2025},
  note   = {R package version 0.8.0},
  url    = {https://CRAN.R-project.org/package=espnscrapeR}
}

@misc{ESPN2025,
  author = {{ESPN}},
  title  = {{ESPN Play-by-Play and Win Probability Data}},
  year   = {2025},
  url    = {https://www.espn.com/},
  note   = {Accessed 2025-12-21}
}

@article{Dawid1982,
  author  = {Dawid, A. P.},
  title   = {{The Well-Calibrated Bayesian}},
  journal = {Journal of the American Statistical Association},
  volume  = {77},
  number  = {379},
  pages   = {605--613},
  year    = {1982},
  doi     = {10.1080/01621459.1982.10477856}
}

@article{GneitingRaftery2007,
  author  = {Gneiting, Tilmann and Balabdaoui, Fadoua and Raftery, Adrian E.},
  title   = {{Probabilistic forecasts, calibration and sharpness}},
  journal = {Journal of the Royal Statistical Society: Series B (Statistical Methodology)},
  volume  = {69},
  number  = {2},
  pages   = {243--268},
  year    = {2007},
  doi     = {10.1111/j.1467-9868.2007.00587.x}
}

@article{FosterVohra1998,
  author  = {Foster, Dean P. and Vohra, Rakesh V.},
  title   = {{Asymptotic Calibration}},
  journal = {Biometrika},
  volume  = {85},
  number  = {2},
  pages   = {379--390},
  year    = {1998},
  doi     = {10.1093/biomet/85.2.379}
}

@article{Brier1950,
  author  = {Brier, Glenn W.},
  title   = {Verification of Forecasts Expressed in Terms of Probability},
  journal = {Monthly Weather Review},
  volume  = {78},
  number  = {1},
  pages   = {1--3},
  year    = {1950},
  doi     = {10.1175/1520-0493(1950)078<0001:VOFEIT>2.0.CO;2}
}

@article{Murphy1973,
  author  = {Murphy, Allan H.},
  title   = {A New Vector Partition of the Probability Score},
  journal = {Journal of Applied Meteorology},
  volume  = {12},
  number  = {4},
  pages   = {595--600},
  year    = {1973},
  doi     = {10.1175/1520-0450(1973)012<0595:ANVPOT>2.0.CO;2}
}

@article{DeGrootFienberg1983,
  author  = {DeGroot, Morris H. and Fienberg, Stephen E.},
  title   = {The Comparison and Evaluation of Forecasters},
  journal = {Journal of the Royal Statistical Society: Series D (The Statistician)},
  volume  = {32},
  number  = {1-2},
  pages   = {12--22},
  year    = {1983},
  doi     = {10.2307/2987588}
}

@article{Dawid1984,
  author  = {Dawid, A. P.},
  title   = {Statistical Theory: The Prequential Approach},
  journal = {Journal of the Royal Statistical Society: Series A (General)},
  volume  = {147},
  number  = {2},
  pages   = {278--292},
  year    = {1984},
  doi     = {10.2307/2981683}
}

@article{RamdasGrunwaldVovkShafer2023,
  author  = {Ramdas, Aaditya and Gr{\"u}nwald, Peter and Vovk, Vladimir and Shafer, Glenn},
  title   = {Game-Theoretic Statistics and Safe Anytime-Valid Inference},
  journal = {Statistical Science},
  volume  = {38},
  number  = {4},
  pages   = {576--597},
  year    = {2023},
  doi     = {10.1214/23-STS894}
}

@article{ArnoldHenziZiegel2023,
  author  = {Arnold, David and Henzi, Alexander and Ziegel, Johanna F.},
  title   = {Sequentially Valid Tests for Forecast Calibration},
  journal = {The Annals of Applied Statistics},
  volume  = {17},
  number  = {3},
  pages   = {1909--1935},
  year    = {2023},
  doi     = {10.1214/22-AOAS1697}
}

@article{Stern1994,
  author  = {Stern, Hal S.},
  title   = {A Brownian Motion Model for the Progress of Sports Scores},
  journal = {Journal of the American Statistical Association},
  volume  = {89},
  number  = {427},
  pages   = {1128--1134},
  year    = {1994},
  doi     = {10.1080/01621459.1994.10476851}
}

@article{LockNettleton2014,
  author  = {Lock, Dennis and Nettleton, Dan},
  title   = {Using Random Forests to Estimate Win Probability Before Each Play of an {NFL} Game},
  journal = {Journal of Quantitative Analysis in Sports},
  volume  = {10},
  number  = {2},
  pages   = {197--205},
  year    = {2014},
  doi     = {10.1515/jqas-2013-0100}
}

@article{YurkoVenturaHorowitz2019,
  author  = {Yurko, Ronald and Ventura, Samuel and Horowitz, Maksim},
  title   = {{nflWAR}: A Reproducible Method for Offensive Player Evaluation in Football},
  journal = {Journal of Quantitative Analysis in Sports},
  volume  = {15},
  number  = {3},
  pages   = {163--183},
  year    = {2019},
  doi     = {10.1515/jqas-2018-0010}
}

@article{YehRiceDubin2022,
  author  = {Yeh, Chi-Kuang and Rice, Gregory and Dubin, Joel A.},
  title   = {Evaluating Real-Time Probabilistic Forecasts With Application to {National Basketball Association} Outcome Prediction},
  journal = {The American Statistician},
  volume  = {76},
  number  = {3},
  pages   = {214--223},
  year    = {2022},
  doi     = {10.1080/00031305.2021.1967781}
}

@article{Owen2007,
  author  = {Owen, Art B.},
  title   = {The Pigeonhole Bootstrap},
  journal = {The Annals of Applied Statistics},
  volume  = {1},
  number  = {2},
  pages   = {386--411},
  year    = {2007},
  doi     = {10.1214/07-AOAS122}
}

@article{DaveziesDHAultfoeuilleGuyonvarch2021,
  author  = {Davezies, Laurent and D'Haultf{\oe}uille, Xavier and Guyonvarch, Yannick},
  title   = {Empirical Process Results for Exchangeable Arrays},
  journal = {The Annals of Statistics},
  volume  = {49},
  number  = {2},
  pages   = {845--862},
  year    = {2021},
  doi     = {10.1214/20-AOS1981}
}

@article{Menzel2021,
  author  = {Menzel, Konrad},
  title   = {Bootstrap With Cluster-Dependence in Two or More Dimensions},
  journal = {Econometrica},
  volume  = {89},
  number  = {5},
  pages   = {2143--2188},
  year    = {2021},
  doi     = {10.3982/ECTA15383}
}

@misc{ESPN2019BullsLakers,
  author = {{ESPN}},
  title  = {{Los Angeles Lakers 118, Chicago Bulls 112}},
  year   = {2019},
  url    = {https://www.espn.com/nba/game/_/gameId/401160742/lakers-bulls},
  note   = {Win-probability path accessed 2026-07-19}
}
\end{document}